\documentclass[journal]{IEEEtran}
\usepackage{amsmath,amsfonts}
\usepackage{amssymb}
\usepackage{algorithm}
\newcommand{\lgamma}{\mathrm{lgamma}}
\usepackage{bm}    
\usepackage{array}
\usepackage[caption=false,font=small,labelfont=sf,textfont=sf]{subfig}
\usepackage{textcomp}
\usepackage[T1]{fontenc}
\usepackage{newtxtext,newtxmath}
\usepackage{microtype}
\usepackage{stfloats}
\usepackage{url}
\usepackage{verbatim}
\usepackage{graphicx}
\usepackage[colorlinks, citecolor=green]{hyperref}
\def\BibTeX{{\rm B\kern-.05em{\sc i\kern-.025em b}\kern-.08em
		T\kern-.1667em\lower.7ex\hbox{E}\kern-.125emX}}
\usepackage{balance}
\usepackage{xcolor}
\usepackage{multirow}
\usepackage{algpseudocode}
\usepackage[table]{xcolor} 
\usepackage{cite}
\usepackage{booktabs}

\begin{document} 
	\title{Unified ROI-based Image Compression Paradigm with Generalized Gaussian Model}
	\author{Kai Hu, 
		Junfu Tan, 
		Fang Xu,
		Ramy Samy,
		Yu Liu
		\thanks{Kai Hu, Junfu Tan, and Yu Liu are with Tianjin University, Tianjin 300072, China (email: kaihu@tju.edu.cn). Fang Xu is with Southwest University, Chongqing, China. Ramy Samy is with the Communication Department, Space Technology Center, Egypt. \textit{Corresponding authors:Yu Liu, Fang Xu} (email: liuyu@tju.edu.cn, 2011204017@tju.edu.cn).}
		\markboth{Journal of \LaTeX\ Class Files,~Vol.~18, No.~9, September~2025}%
		{How to Use the IEEEtran \LaTeX \ Templates}
		
	}
	
	\maketitle
	
	\begin{abstract}
		Region-of-Interest (ROI)-based image compression allocates bits unevenly according to the semantic importance of different regions. Such differentiated coding typically induces a sharp-peaked and heavy-tailed distribution. This distribution characteristic mathematically necessitates a probability model with adaptable shape parameters for accurate description. However, existing methods commonly use a Gaussian model to fit this distribution, resulting in a loss of coding performance. To systematically analyze the impact of this distribution on ROI coding, we develop a unified rate-distortion optimization theoretical paradigm. Building on this paradigm, we propose a novel Generalized Gaussian Model (GGM) to achieve flexible modeling of the latent variables distribution. To support stable optimization of GGM, we introduce effective differentiable functions and further propose a dynamic lower bound to alleviate train-test mismatch. Moreover, finite differences are introduced to solve the gradient computation after GGM fits the distribution. Experiments on COCO2017 demonstrate that our method achieves state-of-the-art in both ROI reconstruction and downstream tasks (e.g., Segmentation, Object Detection). Furthermore, compared to classical probability models, our GGM provides a more precise fit to feature distributions and achieves superior coding performance. The project page is at https://github.com/hukai-tju/ROIGGM.
	\end{abstract}
	

	\section{Introduction}
	\IEEEPARstart{T}{he} digital age has brought an unprecedented surge in high-resolution visual data. Much of this content is internally redundant, and only a small portion is truly critical. Region-of-Interest (ROI) image compression does not allocate bits uniformly like traditional global compression, but allocates more bits to information areas and fewer bits to less important areas under a fixed bit budget. The traditional standard (e.g., JPEG2000) performs this principle using wavelet transforms and the Embedded Block Coding with Optimized Truncation (EBCOT) algorithm, thereby systematizing ROI coding~\cite{tahoces2008image}.
	Similarly, ROI-based neural image compression (NIC) leverages data-driven approaches to achieve more flexible and efficient coding. This paradigm facilitates a shift from \textit{uniform fidelity} to \textit{semantic-driven, fidelity-on-demand} image compression, which is particularly valuable for applications with stringent requirements on local information integrity, such as medical diagnostics, video surveillance and conferencing, autonomous driving perception, and remote sensing monitoring.
	\begin{figure}[!t] 
		\centering    
		\includegraphics[scale=0.375]{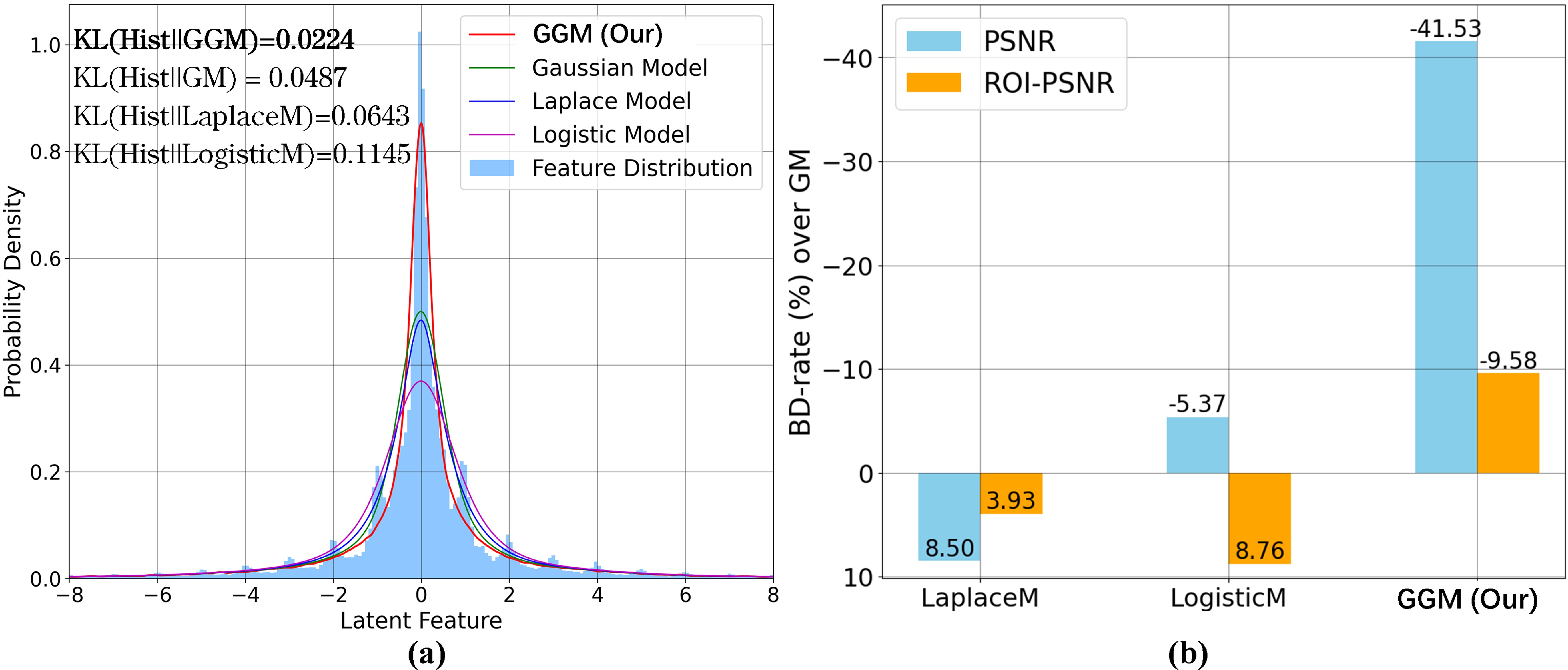} 
		\caption{(a) Comparison of latent feature histograms and fitted distributions, with corresponding Kullback-Leible (KL) divergence values. Our GGM can more accurately fit the peaked and heavy-tailed distribution of the latent variables, achieving the lowest KL divergence of 0.0224. (b) Achieved BD-rate savings compared to the GM baseline for each probability model. For both the entire image and the target region, our GGM achieves optimal coding performance in terms of BD-rate, significantly outperforming classical probability models.}
		\label{fig:histpdfs}
		\vspace{-0.2cm}
	\end{figure}

		In recent years, NIC has rapidly progressed thanks to its superior coding performance~\cite{balle2016end,minnen2018joint,cheng2020learned,liu2023learned,9989403}. The field was significantly advanced by the end-to-end learned frameworks introduced by Ballé et al.~\cite{balle2016end}, which typically comprise an analysis transform, a synthesis transform, a quantizer, and an entropy model. Their subsequent work~\cite{balle2018variational} incorporated a scale hyperprior network, utilizing conditional Gaussian priors to model spatial dependencies. Building on this, Minnen et al.~\cite{minnen2018joint} employed an autoregressive spatial entropy model to leverage causal context for prediction. 
		Along with these key advancements, ROI-based NIC has emerged as an important and rapidly growing research focus~\cite{li2018learning,mentzer2018conditional,cai2018efficient,cai2019end,li2020learning,ma2021variable}. Li et al.~\cite{li2018learning,li2020learning} and Mentzer et al.~\cite{mentzer2018conditional} relied on generating explicit importance maps to guide adaptive bit allocation. Cai et al.~\cite{cai2019end} built upon~\cite{cai2018efficient} by applying ROI masks to multiscale feature representations via dot products.  To address the performance limitations of pre-quantization masking, recent works~\cite{li2024saliency,text2025,hu2025} introduce more sophisticated techniques. Li et al.~\cite{li2024saliency} decomposed latent features into base and enhancement layers prior to quantization, applying saliency-based masks to selectively zero-out background areas in an enhancement layer. Our recent works~\cite{text2025,hu2025} implemented joint implicit bit allocation and dual decoders, achieving high-quality region-adaptive coding without relying on explicit pre-quantization masking.
		
		However, we observe significant spatial heterogeneity in the latent variables distribution of ROI-based compression, where the statistical characteristics of ROI and non-ROI regions appear to be fundamentally distinct. This distribution characteristic urgently requires a distribution with variable shape parameters to describe mathematically. As Fig.~\ref{fig:histpdfs} illustrates, the global latent variables distribution is sharp-peaked and heavy-tailed. The prominent peak near zero predominantly comprises background features, whereas the tails on both sides are almost entirely dominated by ROI features.  Existing ROI-based compression methods~\cite{ma2021variable,li2024saliency,text2025} typically assume a single probability model (e.g., Gaussian). Such a simplified assumption cannot accurately capture the underlying latent variables distribution, which in turn leads to suboptimal entropy coding and degrades overall compression performance. Some NIC methods~\cite{9156817,fu2023learned,10938038} have adopted more flexible prior distributions to move beyond the limitations of a single Gaussian model (GM). Examples include Gaussian Mixture Models (GMM)~\cite{9156817} and Gaussian–Laplacian–Logistic Mixture Models (GLLMM)~\cite{fu2023learned}. Zhang et al.~\cite{10938038} employed the Generalized Gaussian Model (GGM) and designed shape parameters of different complexities to enhance the ability to model the distributions of natural images. However, these approaches are not designed for ROI coding and overlook its inherent spatial heterogeneity. Furthermore, mixture models~\cite{9156817}\cite{fu2023learned} often incur high complexity, and the employed GGM~\cite{10938038} neglects the importance of optimizing its scale and shape parameters for achieving accurate modeling of latent variables distribution.
		
		The latent variables distribution in ROI compression is shaped by the rate-distortion (RD) function and the network architecture. Accordingly, this paper first develops a unified rate-distortion optimization (RDO) theoretical paradigm. This paradigm not only consolidates prior design choices for ROI coding (e.g., implicit/explicit bit allocation strategies~\cite{li2020learning,text2025} and region-differentiated loss), but also focuses on probability models. As an extension of the GM, the GGM~\cite{wang2006fast} provides two approximately orthogonal control parameters—a scale $\mathbf{\alpha}$ and a shape $\mathbf{\beta}$. Building on the above framework, we further propose a novel GGM to fully leverage the flexibility of learned $\mathbf{\alpha}$ and learned $\mathbf{\beta}$ to accurately model complex spatially heterogeneous distributions. We then design specialized and differentiable activation functions tailored to each parameter's properties: a \textit{Softplus} function for $\mathbf{\beta}$ and a \textit{Huber-like} function for $\mathbf{\alpha}$. They not only facilitate stable optimization of the GGM but also incorporate meaningful information to enhance the accuracy of latent distributions modeling. A train-test mismatch primarily stems from the discrepancy between additive uniform noise during training and hard rounding during testing.  Furthermore, we note that small $\mathbf{\alpha}$ values exacerbate a train-test mismatch and degrade coding performance~\cite{zhang2023uniform}. To mitigate this, we introduce a dynamic lower bound strategy for $\mathbf{\alpha}$. In addition, we address the analytical intractability of the gradient computation for the regularized lower incomplete gamma function $P(a,b)$ within the GGM~\cite{blahak2010efficient} by introducing an efficient finite central difference scheme, thereby enabling stable end-to-end optimization.

		Our main contributions are summarized as follows: 
		\begin{itemize}
			\item{We formulate a unified RDO theoretical paradigm for ROI coding. This paradigm focuses on the core bit allocation strategies, probability models, and RD functions.}
			\item{Based on the above paradigm, we introduce an elaborate GGM  for ROI-based image compression. Experimental results demonstrate that our GGM provides more precise fitting of latent variables distribution compared to classical probability models—namely, Gaussian, Laplacian, and Logistic models and their mixture models—while achieving superior coding performance.}
			\item{To address stability optimization during GGM training, we design effective and differentiable activation functions for the learned scale $\mathbf{\alpha}$ and the learned shape $\mathbf{\beta}$, respectively. Furthermore, we propose a dynamic lower bound to mitigate the train-test mismatch for the scale $\mathbf{\alpha}$.}
			\item{To resolve the complex numerical integration in gradient calculation of the $P(a,b)$ function within the GGM, we design an efficient and robust gradient approximation estimation scheme to achieve end-to-end stable optimization.}
		\end{itemize}
		
		The remainder of this paper is organized as follows. Section~\ref{2} provides a brief review of related work on ROI compression and probability models for image compression. Section~\ref{3} establishes an RDO theoretical paradigm for ROI coding. Section~\ref{4} provides a detailed description of our GGM. Finally, Sections~\ref{5} and~\ref{6} present the experimental results and conclusions, respectively.

		\section{Related Work}
		\label{2}
		\subsection{ROI-based Image Compression}
		Tahoces et al.~\cite{tahoces2008image} adopted a selective coefficient mask shift coding method for the JPEG2000 standard to selectively offset the ROI coefficients of different resolution sub-bands, achieving fine control over the relative quality between ROI and background. 
		With the advancement of convolutional neural networks (CNNs), ROI-based data-driven image compression has become a new focus. Li et al.~\cite{li2018learning, li2020learning} introduced a mapping subnet for key regions to generate explicit important mappings, which adaptively allocate bits based on image contents. Mentzer et al.~\cite{mentzer2018conditional} also used importance maps to guide bit allocation. Based on~\cite{cai2018efficient}, Cai et al.~\cite{cai2019end} used ROI masks to dot-product multi-scale representations, which mainly allows more bits to be allocated to the foreground in the image.  Ma et al.~\cite{ma2021variable} adopted 3D masks to prune latent representations to improve bit allocation in background regions and employed GANs for supervised training of the entire coding network. Li et al.~\cite{li2023roi} integrated ROI masks into different layers of the compression network, achieving spatial adaptability and variable-rate compression by modifying the Lagrange multiplier $\lambda$ in different regions. Li et al.~\cite{li2024saliency} developed a saliency segmentation-oriented deep compression model, employing a saliency segmentation model to generate masks. For bit allocation across different image regions, they decomposed latent features into base and enhancement layers before quantization, applying masks to locally zero-out background areas in the enhancement layer, followed by entropy coding using a Double-Scale Entropy Module. 
		
		Peng et al.~\cite{peng2024saliency} adopted a saliency map-guided image compression network for machine-vision coding tasks. Jin et al.~\cite{jin2025customizable} introduced a customizable ROI-based deep image compression paradigm incorporating user-provided textual information. Their proposed framework enables both customizable ROI compression and user-adjustable quality trade-offs between ROI and non-ROI reconstruction. Hu et al.~\cite{text2025,hu2025} applied implicit bit allocation and  dual decoders to ROI-based NIC models, achieving high-quality region-adaptive coding.  It also confirms that implicit methods are more flexible and perform better than explicit bit allocation.
		
		\subsection{Probabilistic Distribution Models for Image Compression}
		Probability distribution models are foundational for accurately predicting latent feature distributions, which is critical for achieving efficient compression. Ballé et al.~\cite{balle2016end} used piece-wise linear functions to non-parametric probability model the edge distribution of each quantized latent tensor, avoiding model hypothesis bias caused by artificial priors. Ballé et al.~\cite{balle2018variational} introduced scale hyperpriors to model latent tensors as conditional Gaussian scale models (GSM), which can effectively model the spatial dependencies between latent tensors. Minnen et al.~\cite{minnen2018joint} extended the GSM to a GMM, while predicting the mean and scale parameters of each latent tensor. An autoregressive model is then used to model the causal context of the latent tensor to improve coding performance. Cheng et al.~\cite{cheng2020learned} used a multi-group partitioned GMM to model the distribution of latent tensors instead of a single Gaussian distribution, providing more refined probability modeling. Each latent tensor is composed of multiple weighted Gaussian components, which can flexibly model complex distributions such as multimodal and asymmetric. A more flexible mixture model, combining Gaussian, Laplacian, and logistic distributions, was proposed by Fu et al.~\cite{fu2023learned} to accurately fit the feature distributions of all images or of different regions within an image.
		
		Peng et al.~\cite{peng2025generalized} used a Generalized Gaussian distribution for point cloud attribute compression and controlled the distribution with shape parameter $\mathbf{\beta}$ to fit the latent tensor more accurately. Dynamic likelihood intervals are also used to improve the problems of existing methods for arithmetic encoding. To achieve a better balance between compression performance and complexity, Zhang et al.~\cite{zhang2025generalized} extended the GM with mean and scale parameters to the GGM. They also designed shape parameters of different complexities and adopted some improved training methods to achieve high-quality coding of natural images. Jiang et al.~\cite{jiang2025omr} used different probability models to accurately model the distribution of high-frequency and low-frequency components in screen content images.
		
		\section{ROI Coding Paradigm}
		\label{3}
		In this section, we systematically analyze the RDO for ROI-based image compression. Based on the feature distribution characteristics of ROI coding, we formulate a unified theoretical paradigm built upon both the GGM.
		
		\subsection{RDO under Spatial Heterogeneity Constraints}
		The distribution of latent features in ROI coding is not intrinsic, it is jointly shaped by RD objective functions and network architectures~\cite{cai2019end,li2020learning,text2025}.
		Given a binary importance map \(\mathbf{M}\in\{0,1\}^{H\times W}\) generated by a CNN for an input image $\mathbf{x}$, where \(M_i = 1\) indicates pixel \(i\) belongs to the ROI, we apply distinct penalty factors to the reconstruction errors in ROI and non-ROI regions. A straightforward yet effective weighting strategy is \(
		w_i = \left\{
		\begin{aligned}
			w_{\text{roi}}, \quad     M_i = 1. \\
			w_{\text{nonroi}}, \quad  M_i = 0.
		\end{aligned}
		\right.\)
		With the constraint \( w_{\text{roi}} > w_{\text{nonroi}} \), this weighting enables differentiated content reconstruction. Let $\mathbf{\phi}$ and $\mathbf{\theta}$ denote the parameters of the conditional encoder $g_a(\mathbf{x},\mathcal{F}(\mathbf{M}); \mathbf{\phi})$ and the decoder $g_s(\hat{\mathbf{y}};\mathbf{\theta})$, respectively. Here, $\mathcal{F}(\mathbf{M})$ is a transform applied to the importance map $\mathbf{M}$, which implements specific bit allocation strategies (e.g., implicit/explicit bit allocation~\cite{cai2019end,li2020learning,text2025}) for latent features $\mathbf{y}$ within the encoder.
		
		Under these constraints, the encoder $g_a$ is driven to generate high-variance, information-rich latent features in ROI regions to reduce weighted distortion, while producing low-variance, near-zero features in background regions to lower the bitrate, as shown in Fig.~\ref{fig:histpdfs}. This spatially differentiated optimization results in a global latent feature distribution that exhibits pronounced spatial heterogeneity as well as sharp-peaked and heavy-tailed characteristics. The standard variational autoencoder (VAE)-based compression framework typically assumes that the posterior distribution of latent variables \( q_\phi(\hat{\mathbf{y}}|\mathbf{x}) \) can be approximated by a parameterized Gaussian prior \( p_\theta(\hat{\mathbf{y}}) \sim \mathcal{N}(\mathbf{\mu}, \mathbf{\sigma}^2) \)~\cite{he2022elic,10210338,10735255}.  
		However, under the influence of ROI-weighted distortion, the true posterior distribution \( q_\phi(\hat{\mathbf{y}}|\mathbf{x}) \) undergoes severe shape deformation~\cite{text2025}. Consequently, directly employing a Gaussian prior results in an irreducible lower bound in the KL divergence—that is, a theoretical variational gap. For instance, as shown in Fig.~\ref{fig:histpdfs}(a), using a GM results in a KL divergence of 0.0487, whereas our GGM achieves a KL divergence of only 0.0224.

		\subsection{Unified Rate-Distortion Optimization paradigm}
		To bridge this variational gap, we introduce the GGM as a prior and integrate it with RD functions and bit allocation strategies, thereby establishing a unified RDO theoretical paradigm for ROI compression. According to~\cite{balle2018variational,bai2024deep}, universal RDO functions can be expressed by introducing a Lagrange multiplier $\lambda$:
		\begin{equation}
			\begin{aligned}
				\mathcal{L}(\mathbf{\phi},\mathbf{\theta}) = \underbrace{\mathbb{E}_{q_\phi(\hat{\mathbf{y}}|\mathbf{x})}[- \log p_\theta(\mathbf{x}|\hat{\mathbf{y}})]}_{Distortion} + \lambda \cdot \underbrace{D_{KL}[q_\phi(\hat{\mathbf{y}}|\mathbf{x}) || p_\theta(\hat{\mathbf{y}})]}_{Rate},
			\end{aligned}
			\label{eq4}
		\end{equation}
		where  $p_{\theta}(\mathbf{x}|\hat{\mathbf{y}})$ represents the likelihood (decoder $g_s$ output distribution) and $p_{\theta}(\hat{\mathbf{y}})$ represents the prior probability model. $D_{KL}[\cdot||\cdot]$ denotes the KL divergence, and $\hat{\mathbf{y}}$ is the quantized result. We seek to derive both the Distortion and Rate terms using the GGM to handle the distinct statistical properties of ROI coding. 
		The distortion term corresponds to the negative log-likelihood of the decoding results given the latents. To account for the potential distribution characteristics in ROI coding, we model the decoder's output distribution $p_{\theta}(\mathbf{x}|\hat{\mathbf{y}})$ as a GGM with mean $g_s(\hat{\mathbf{y}};\mathbf{\theta})$, scale $\mathbf{\alpha}'$, and shape $\mathbf{\beta}'$:
		\begin{equation}
			p_\theta(\mathbf{x}|\hat{\mathbf{y}}) = \frac{\mathbf{\beta}^{'}}{2 \mathbf{\alpha}^{'} \,\Gamma(1/{\mathbf{\beta}^{'}})}
			\exp\Bigl(-\bigl|\tfrac{\mathbf{x} - g_s(\hat{\mathbf{y}};\mathbf{\theta}))}{\mathbf{\alpha}^{'}}\bigr|^{\mathbf{\beta}^{'}}\Bigr), 
		\end{equation}
		where $\Gamma(\cdot)$ is a gamma function. Specifically, we assume that $\alpha^{'}$ and $\beta^{'}$ are constants. The distortion term $\mathcal{D}$ in Eq. \eqref{eq4} is defined as:
		\begin{equation}
			\begin{aligned}
				\mathcal{D} &= \mathbb{E}_{q_\phi(\hat{\mathbf{y}}\mid \mathbf{x},\mathcal{F}(M))}\Big[-\log\frac{\beta^{'}}{2\alpha^{'}\,\Gamma(1/{\beta^{'}})} + \Bigl|\tfrac{\mathbf{x} - g_{s}(\hat{\mathbf{y}};\mathbf{\theta})}{\alpha^{'}}\Bigr|^{\beta^{'}}\Big] \\
				&= \mathbb{E}_{q_\phi(\hat{\mathbf{y}}\mid \mathbf{x},\mathcal{F}(\mathbf{M}))}\Big[\frac{|\mathbf{x}-g_{s}(\hat{\mathbf{y}};\mathbf{\theta})|^{\beta^{'}}}{{\alpha^{'}}^{\beta^{'}}}\Big] + \log\!\frac{2\alpha^{'}\Gamma(1/{\beta^{'}})}{\beta^{'}},
			\end{aligned}
		\end{equation} 
		The second term can thus be treated as a constant during optimization. Furthermore, the factor ${\alpha^{'}}^{-\beta^{'}}$ can be absorbed into the Lagrange multiplier $\lambda$. Consequently, the minimized distortion  $\mathcal{D}$ for ROI coding is approximately:
		\begin{equation}
			\mathcal{D} = \mathbb{E}_{q_\phi(\hat{\mathbf{y}}\mid \mathbf{x},\mathcal{F}(\mathbf{M}))}\Big[\sum_{i=1}^N w_i\,|x_i - g_{i,s}(\hat{\mathbf{y}};\mathbf{\theta})|^{\beta^{'}}\Big],
		\end{equation}
		The complete Eq. \eqref{eq4} can then be expressed as
		\begin{equation}
			\begin{aligned}
				\mathcal{L}(\mathbf{\phi},\mathbf{\theta}) &= \mathbb{E}_{q_\phi(\hat{\mathbf{y}}\mid \mathbf{x},\mathcal{F}(\mathbf{M}))}\Big[\sum_{i=1}^N w_i \,|x_i - g_{i,s}(\hat{\mathbf{y}};\mathbf{\theta})|^{\beta^{'}}\Big] \\
				& + \lambda \cdot \mathbb{E}_{q_\phi(\hat{\mathbf{y}}\mid \mathbf{x},\mathcal{F}(\mathbf{M}))}\Big[ \log q_\phi(\hat{\mathbf{y}}|\mathbf{x},\mathcal{F}(\mathbf{M})) - \log p_\theta(\hat{\mathbf{y}}) \Big],
			\end{aligned}
		\end{equation}
		
		\begin{figure}[!t] 
			\centering    
			\includegraphics[scale=0.34]{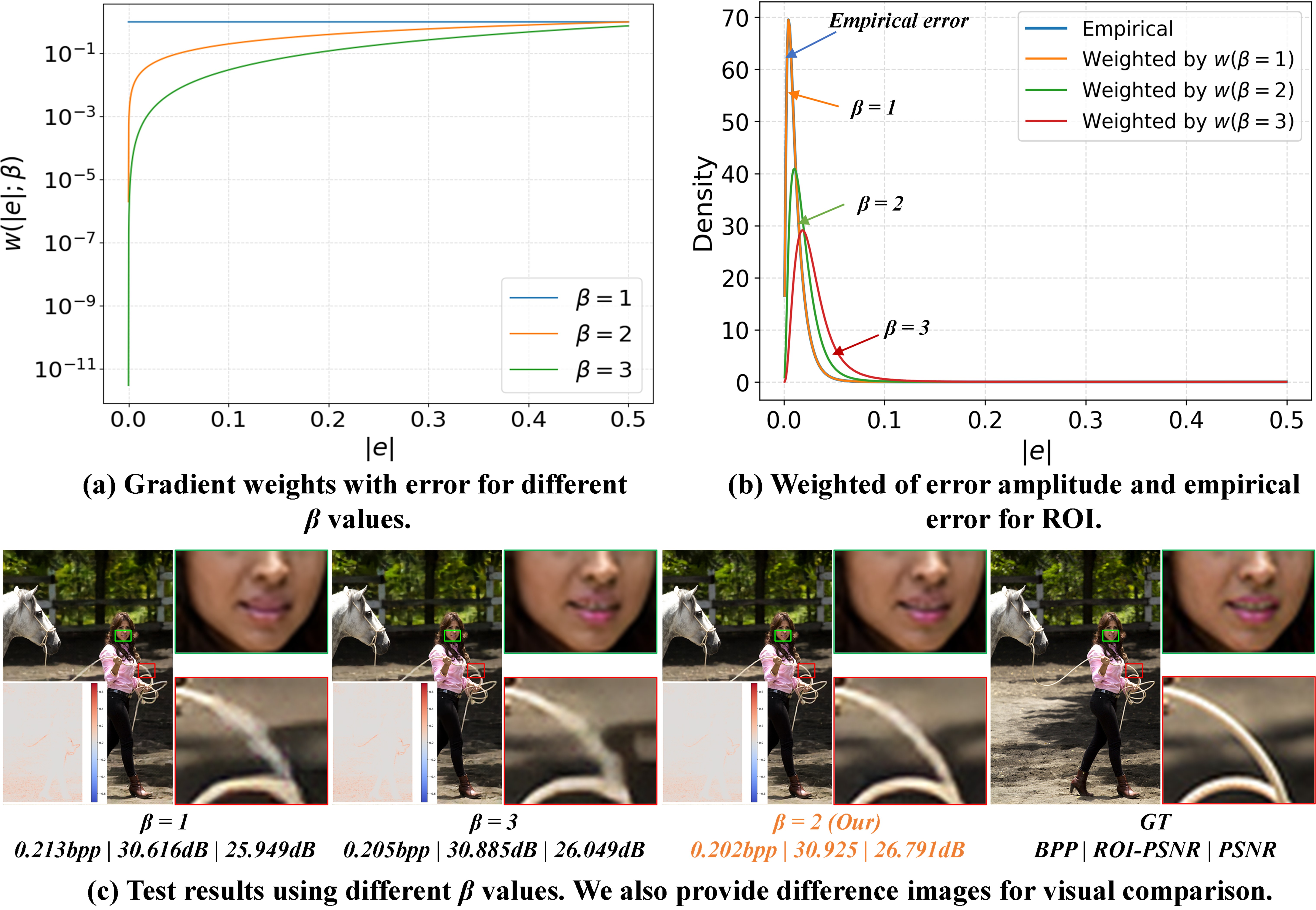} 
			\caption{Analysis of the shape parameter $\beta$. (a) The effective gradient weight $\omega (\lvert e \rvert; \beta) = \beta \lvert e \rvert^{\beta - 1}$ for the distortion term $\mathcal{D}=\sum_i \omega_i \lvert e_i \rvert^{\beta}$. Setting $\beta=1$ gives constant weight; 
				$\beta=2$ yields linear scaling; $\beta=3$ causes quadratic amplification of large errors. (b) Reconstruction error spectra on COCO2017. $\beta=1$ under-penalizes large errors, leading to heavy-tailed residuals; $\beta=3$ over-focuses on outliers; $\beta=2$ achieves the best balance. (c) Rate-distortion performance. The model with $\beta=2$ achieves superior reconstruction quality at lower bitrates.}
			\label{fig:betadist}
			\vspace{-0.15cm}
		\end{figure}
		Guided by the maximum entropy distributions principle~\cite{cover1999elements}—\textit{which states that the Gaussian distribution maximizes entropy under fixed mean and variance}—we set $\beta^{'}=2$ in the GGM for the distortion term $\mathcal{D}$. We also experiment with $\beta^{'}=1, 2$, and $3$ for analysis (Fig.~\ref{fig:betadist}), finding $\beta^{'}=2$ delivers the optimal RD performance. According to~\cite{balle2018variational,9156817}, the Rate term relaxes quantization by adding uniform noise $\mathit{U}(-\frac{1}{2}, \frac{1}{2})$, leading to $q_\phi(\hat{\mathbf{y}} \mid \mathbf{x}, \mathcal{F}(\mathbf{M})) = \prod_i \mathit{U}(\hat{\mathbf{y}}_i|\mathbf{y}_i - \frac{1}{2}, \mathbf{y}_i + \frac{1}{2})$. Consequently, $\log q_\phi(\hat{\mathbf{y}}\mid \mathbf{x},\mathcal{F}(\mathbf{M}))$ is constant and can be dropped.

		Therefore, we derive the final unified RDO coding paradigm based on the GGM as:
		\begin{equation}
			\begin{aligned}
				& p_\theta(\hat{\mathbf{y}}) = \prod_i \Big[\mathcal{N}_{\beta}(\mathbf{\mu}_{\hat{\mathbf{y}},i}, \mathbf{\alpha}^{\mathbf{\beta}}_{\hat{\mathbf{y}},i})*\mathit{U}(- \frac{1}{2}, \frac{1}{2})\Big] (\hat{\mathbf{y}}_i),\\
				&\mathcal{L}(\mathbf{\phi},\mathbf{\theta}) = \underbrace{\mathbb{E}_{q_\phi(\hat{\mathbf{y}}\mid \mathbf{x},\mathcal{F}(\mathbf{M}))} \Big[ \sum_{i=1}^N w_i \,\lVert x_i - g_{i,s}(\hat{\mathbf{y}};\mathbf{\theta})\rVert^2 \Big]}_{Distortion} \\
				& - \lambda \cdot \underbrace{\mathbb{E}_{q_\phi(\hat{\mathbf{y}}\mid \mathbf{x},\mathcal{F}(\mathbf{M}))} \Big[\log p_\theta(\hat{\mathbf{y}}) \Big]}_{Rate},
			\end{aligned}
			\label{eq7}
		\end{equation}
		This framework integrates common bit allocation and region-adaptive losses under the GGM, enabling superior ROI coding.

		\begin{figure}[!t] 
			\centering    
			\includegraphics[scale=0.38]{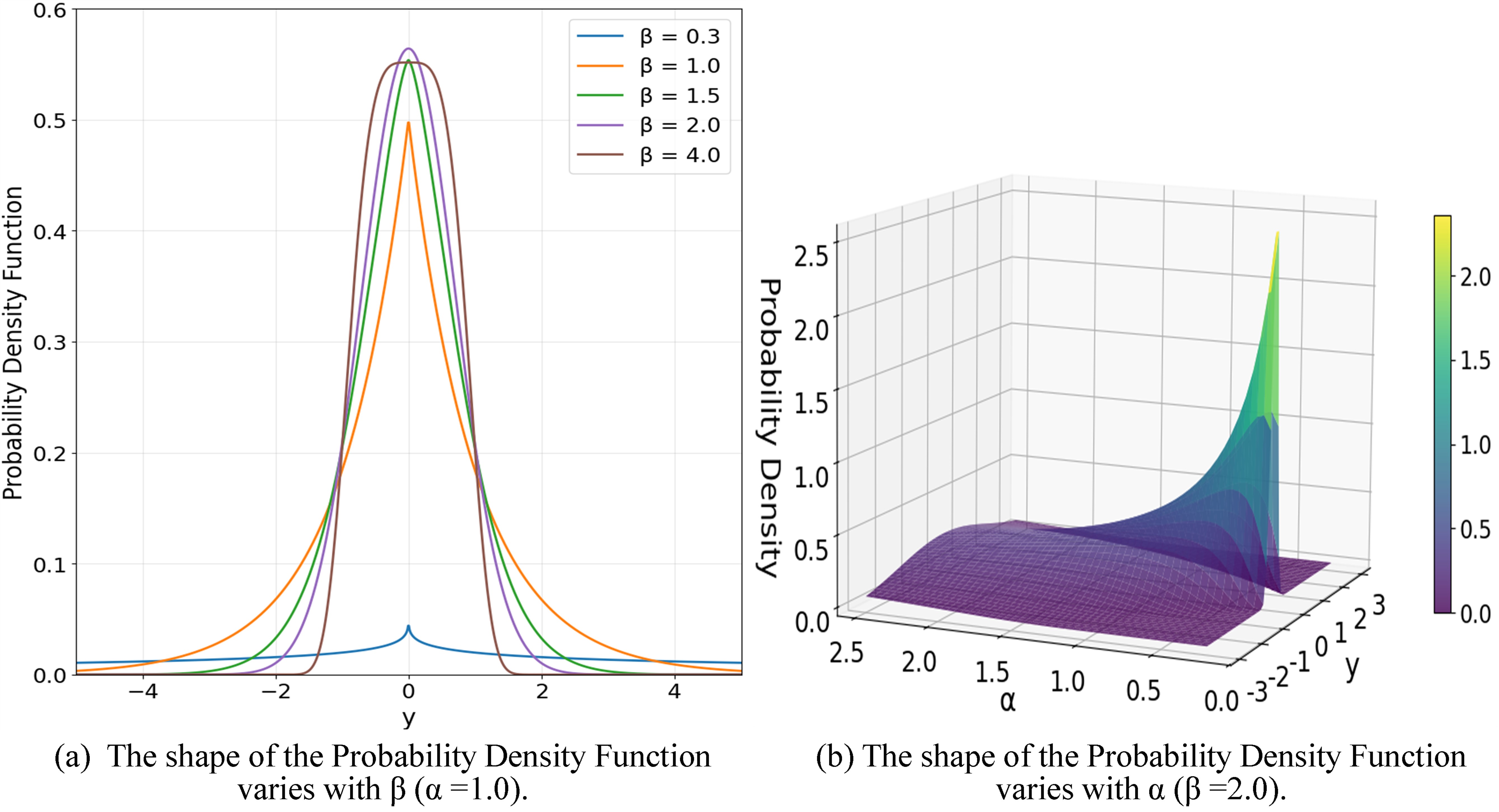} 
			\caption{The influence of the scale ($\alpha$) and shape ($\beta$) parameters on the PDF of the GGM with a fixed mean ($\mu = 0$).}
			\label{fig:pdfshape}
		\end{figure}
		\section{Our Methodology}
		\label{4}
		\subsection{Classic Generalized Gaussian Model}
		The GGM, denoted as $\mathbf{Y} \sim \mathcal{N}_{\beta}(\mathbf{\mu},\mathbf{\alpha}^{\mathbf{\beta}})$, extends the standard GM by introducing a shape parameter $\beta$. This additional parameter allows the GGM to adaptively model distributions with varying tail behaviors, making it particularly suitable for capturing the complex statistics of latent tensors, as shown in Fig.~\ref{fig:pdfshape}. The probability density function
		(PDF) of the symmetric GGM is given by~\cite{wang2006fast,dytso2018analytical} $ f_\beta(\mathbf{y}) = \frac{\mathbf{\beta}}{2\alpha\Gamma(1/\mathbf{\beta})} \exp\left(-\Big(\frac{|\mathbf{y}-\mathbf{\mu}|}{\alpha}\right)^{\mathbf{\beta}}\Big)$. The standard form of the PDF and its corresponding cumulative distribution function (CDF), with $\mu = 0$ and $\alpha = 1$, are:
		\begin{figure}[!t] 
			\centering    
			\includegraphics[scale=0.38]{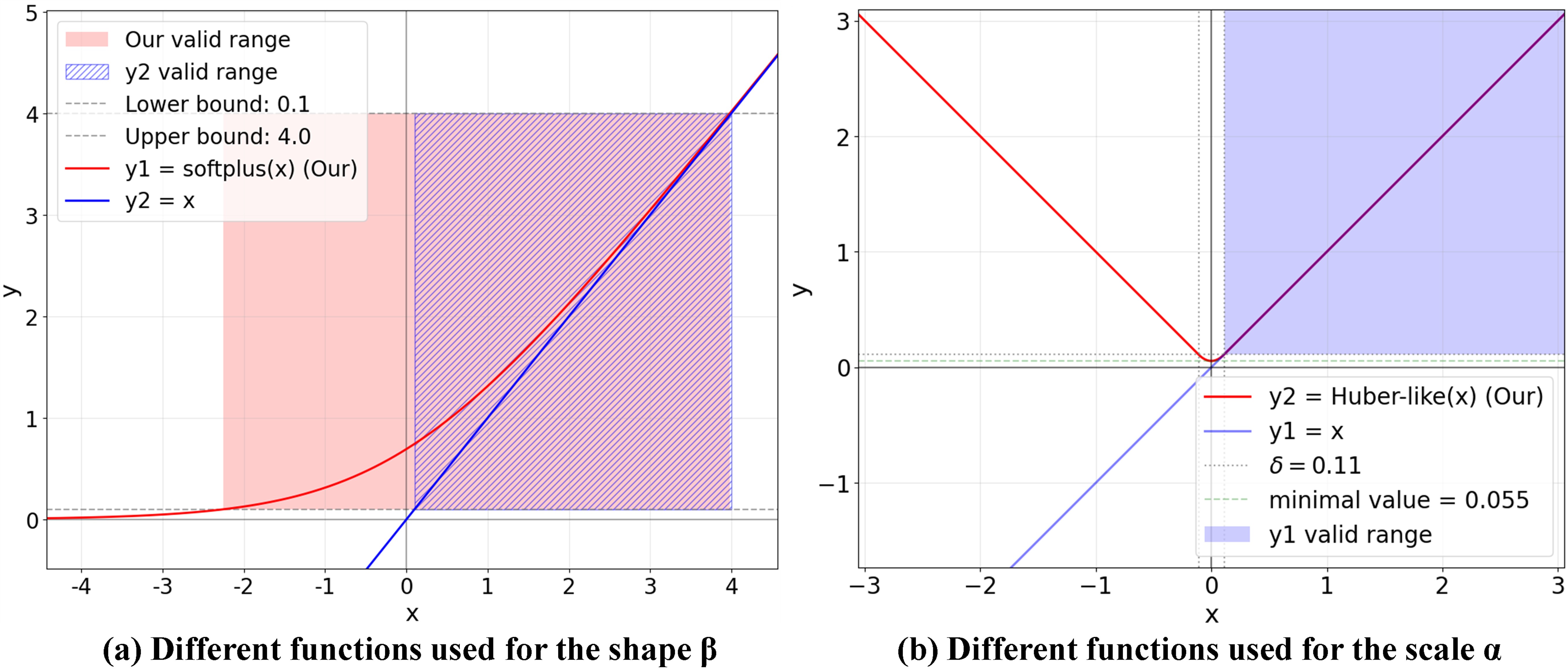} 
			\caption{The functions used for scale $\alpha$ and shape $\beta$. Lower bound values for shape $\beta$ and scale $\alpha$ need to be positive. It is known that Our functions are differentiable within the domain and have a wider effective input range.}
			\label{fig:Func}
			\vspace{-0.15cm}
		\end{figure}
		\begin{equation}
			\begin{aligned}
				& f_\beta(\mathbf{y}) = \frac{\mathbf{\beta}}{2\alpha\Gamma(1/\mathbf{\beta})} \exp\left(-|\mathbf{y}|^{\mathbf{\beta}}\right), \\
				& c_\beta(\mathbf{y}) = \int_{-\infty}^\mathbf{y} f_\beta(u) du = \frac{1}{2} + \frac{\operatorname{sgn}(\mathbf{y})}{2}
				P\left(\frac{1}{\mathbf{\beta}}, |y|^{\mathbf{\beta}}\right),  \\
			\end{aligned}
		\end{equation}
		where $c_\beta(\mathbf{y})$ is the CDF of the GGM, and $P (\cdot, \cdot)$ denotes the regularized lower incomplete gamma function~\cite{blahak2010efficient}. The incomplete gamma function $\gamma(a,b)$ and the Gamma function $\Gamma(a)$ are defined as  
		\begin{equation}
			\begin{aligned}
				& P(a,b) = \frac{\gamma(a,b)}{\Gamma(a)}, \\
				& \Gamma(a) = \int_0^\infty t^{a-1} e^{-t} dt;  \gamma(a,b) = \int_0^b t^{a-1} e^{-t} dt,
			\end{aligned}
			\label{eq5}
		\end{equation}
		Here, we set $a= \frac{1}{\mathbf{\beta}}$ and $b=|\mathbf{y}|^{\mathbf{\beta}}$. The GGM is characterized by two approximately orthogonal learned parameters: the scale $\mathbf{\alpha}$ and the shape $\mathbf{\beta}$. This dual-parameter formulation is its key distinction from single-parameter models like the Gaussian and Laplacian. The scale $\mathbf{\alpha}$ controls the dispersion (width) of the distribution and directly influences entropy: a smaller $\mathbf{\alpha}$ indicates higher predictability and lower bitrate, while a larger $\mathbf{\alpha}$ implies greater uncertainty and higher bitrate. The shape parameter $\mathbf{\beta}$ governs the kurtosis, affecting both peakedness and tail thickness. Empirically, regions with complex textures are better modeled with a smaller $\mathbf{\beta}$, while smoother regions correspond to a larger $\mathbf{\beta}$. Notably, the GGM includes the Gaussian ($\beta=2$, with $\sigma = \alpha/\sqrt{2}$) and Laplacian ($\beta=1$) distributions as special cases.
		\begin{figure}[!t]
			\centering
			\captionsetup[subfigure]{font=scriptsize}
			\subfloat[Feature maps of shape parameter $\beta$ using different functions.]
			{
				\label{fig:betavisual}
				\includegraphics[scale=0.335]{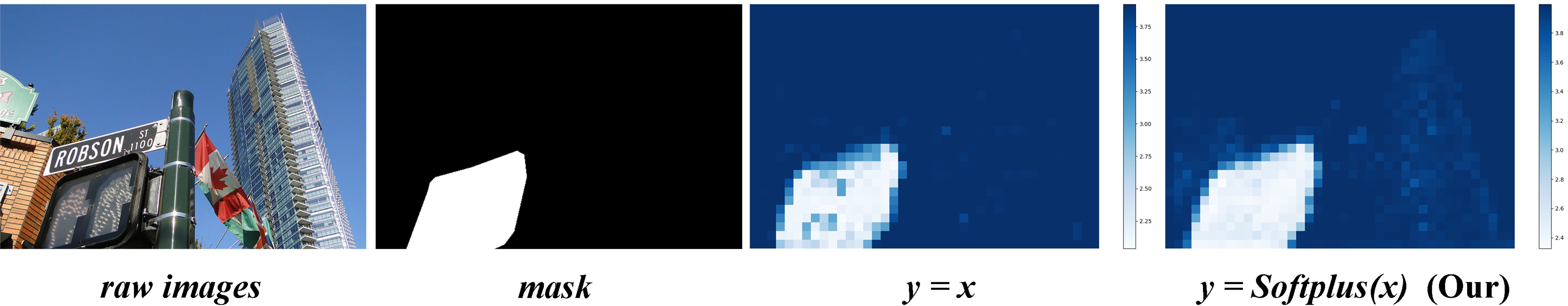} 
			}
			\quad
			\subfloat[Feature maps of scale parameters $\alpha$ using different functions.]
			{
				\label{fig:alphavisual}
				\includegraphics[scale=0.378]{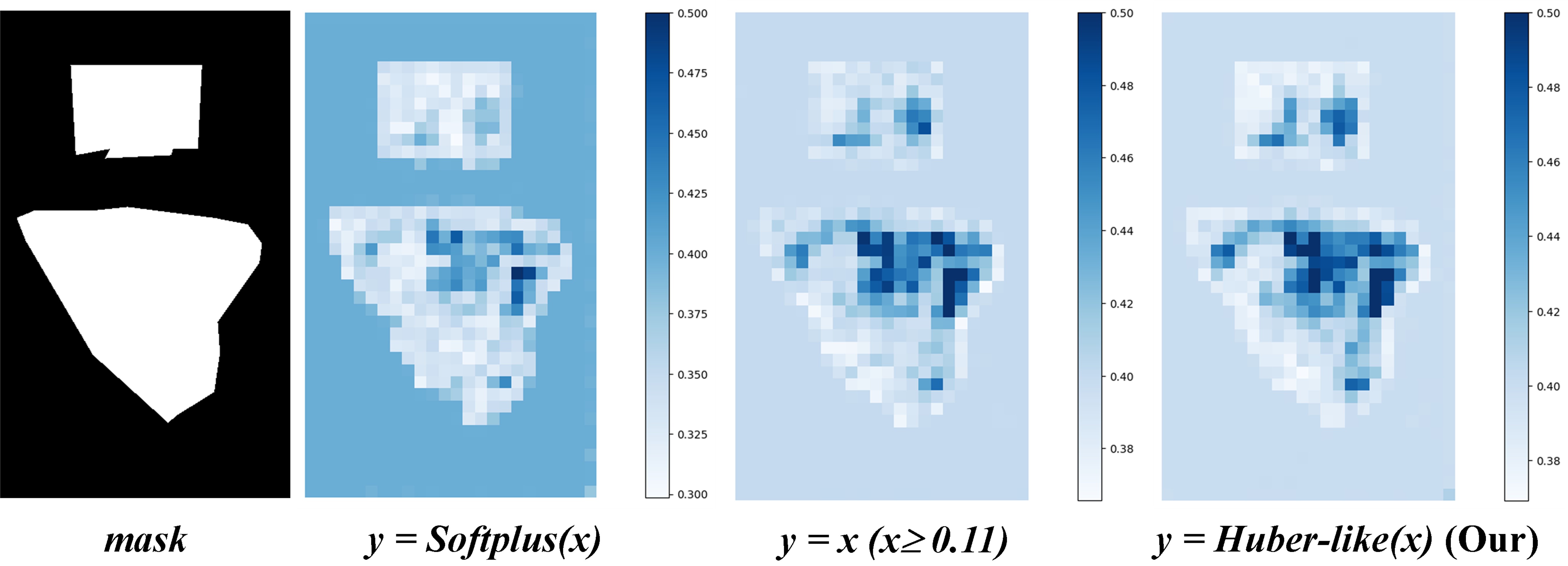} 
			}
			\caption{Feature maps with different functions for shape $\beta$ and scale $\alpha$.}
			\label{fig:alphabeta}
		\end{figure}
		
		\subsection{Our Probability Distribution Model}
		As evidenced in Fig.~\ref{fig:histpdfs}, ROI-based compression results in a spatially heterogeneous latent distribution. Latent features from ROI regions (rich in detail) tend to follow peaked, heavy-tailed distributions, whereas those from non-ROI regions (smoother) are more concentrated and light-tailed. This distribution characteristic mathematically necessitates a distribution with mutable shape parameters for accurate description. To model this heterogeneity, we introduce an elaborate GGM with learned scale ($\mathbf{\alpha}$) and shape ($\mathbf{\beta}$) parameters, enabling precise distribution fitting. However, the stable and effective optimization of $\mathbf{\alpha}$ and $\mathbf{\beta}$ is non-trivial and critical for achieving high-quality coding. To address this challenge, we carefully design optimization strategies for $\mathbf{\alpha}$ and $\mathbf{\beta}$.
		\begin{figure*}[htbp]
			\centering
			\includegraphics[scale=0.535]{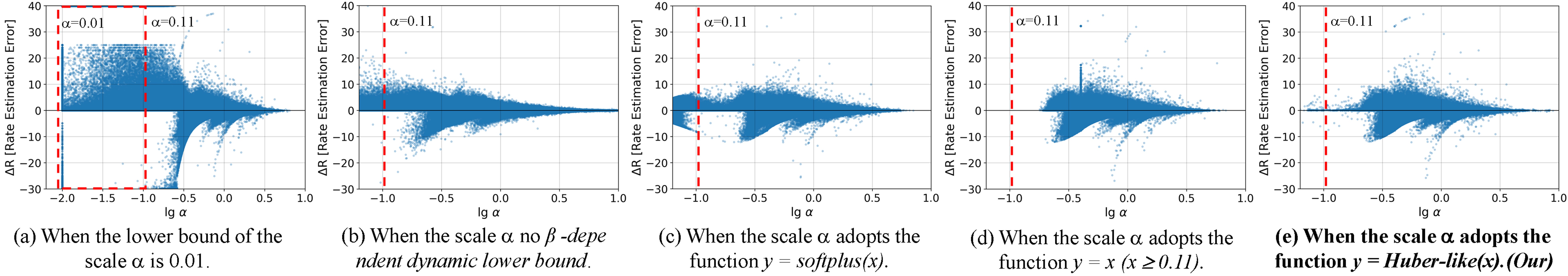} 
			\caption{The rate estimation error $\Delta R =R_{train} - R_{test}$ as a function of the scale $\mathbf{\alpha}$, computed from 100 images and ~$8\times10^6$ samples (Pascal VOC2012~\cite{voc2012}). While $\Delta R$ is generally small and centered near zero, it becomes large and positive when $\mathbf{\alpha}$ is very small (e.g., $<0.11$), indicating severe training-time bitrate overestimation.}
			\label{fig:alphas_r}
		\end{figure*}
		
		\textbf{Shape parameter $\mathbf{\beta}$:} To ensure the shape parameter $\mathbf{\beta}$ remains positive and the overall model differentiable, we introduce a \textit{Softplus} function.
		\begin{equation}
			\begin{aligned}
				f(\mathbf{\beta}) = \textit{Softplus}(\mathbf{\beta}) = \log(1 + e^\mathbf{\beta}) 
			\end{aligned}
		\end{equation}
		The \textit{Softplus} function maps inputs to the positive domain (0, $\infty$). Unlike a linear function \textit{y = x}, its derivative is a \textit{Sigmoid} function, ensuring smooth and well-behaved gradients. As shown in Fig.~\ref{fig:Func}, the \textit{Softplus} function significantly widens the effective range of learnable $\mathbf{\beta}$ values while preventing gradient explosion or vanishing during optimization, thereby facilitating stable training and the learning of meaningful distribution shapes. Furthermore, following~\cite{zhang2025generalized}, we constrain the output of the \textit{Softplus} function to the interval [0.1, 4]. This practical bound allows $\mathbf{\beta}$ to better capture the sparsity and peakedness characteristics typical of latent feature distributions in our task. The effectiveness of our function is visualized in Fig.~\ref{fig:betavisual}, which compares the resulting feature maps of $\mathbf{\beta}$ obtained using different functions. Compared to the simple linear function, our \textit{Softplus} function more accurately delineates high-frequency information in both foreground and background regions, leading to enhanced accuracy in modeling the latent space.
		
		\textbf{Scale parameter $\mathbf{\alpha}$:} To ensure the scale parameter $\mathbf{\alpha}$ remains positive and differentiable, and avoids excessively small values, we design a \textit{Huber-like} activation function, inspired by the Huber loss~\cite{gokcesu2021generalized}. 
		\[ 
		f(\mathbf{\alpha}) = \begin{cases} \frac{\mathbf{\alpha}^2}{2\delta} + \frac{\delta}{2} & \text{if } |\mathbf{\alpha}| \le \delta \\
			\ |\mathbf{\alpha}| & \text{if } |\mathbf{\alpha}| > \delta \end{cases} 
		\]
		Following~\cite{he2022elic,zou2022devil,zhang2023uniform}, we set $\delta=0.11$. This function behaves quadratically for small $|\mathbf{\alpha}|$ and linearly for large $|\mathbf{\alpha}|$ (see Fig.~\ref{fig:Func}), providing stable gradients across the full parameter range. In contrast, the linear function \textit{y = x} (used for the GM in prior work~\cite{he2022elic,zou2022devil,zhang2023uniform}) lacks this adaptive behavior. The scale $\mathbf{\alpha}$ is highly sensitive near zero, as evidenced in Fig.~\ref{fig:alphas_r}. While a \textit{Softplus} function tends to drive small $\mathbf{\alpha}$ values toward zero during training, exacerbating train-test mismatch, our \textit{Huber-like} function effectively mitigates this issue and enhances optimization stability. Furthermore, Fig.~\ref{fig:alphavisual} visualizes the feature maps of $\mathbf{\alpha}$ produced by different functions. We observe that the \textit{Softplus} function over-emphasizes very small values (e.g., in background regions), thereby neglecting the importance of target areas. Conversely, the linear function $\textit{y = x}$ fails to fully map critical target regions. Our function effectively balances the importance of foreground and background while ensuring complete mapping of the target region, enabling accurate modeling of the latent space distribution.
		
		\begin{figure}[!t] 
			\centering    
			\includegraphics[scale=0.37]{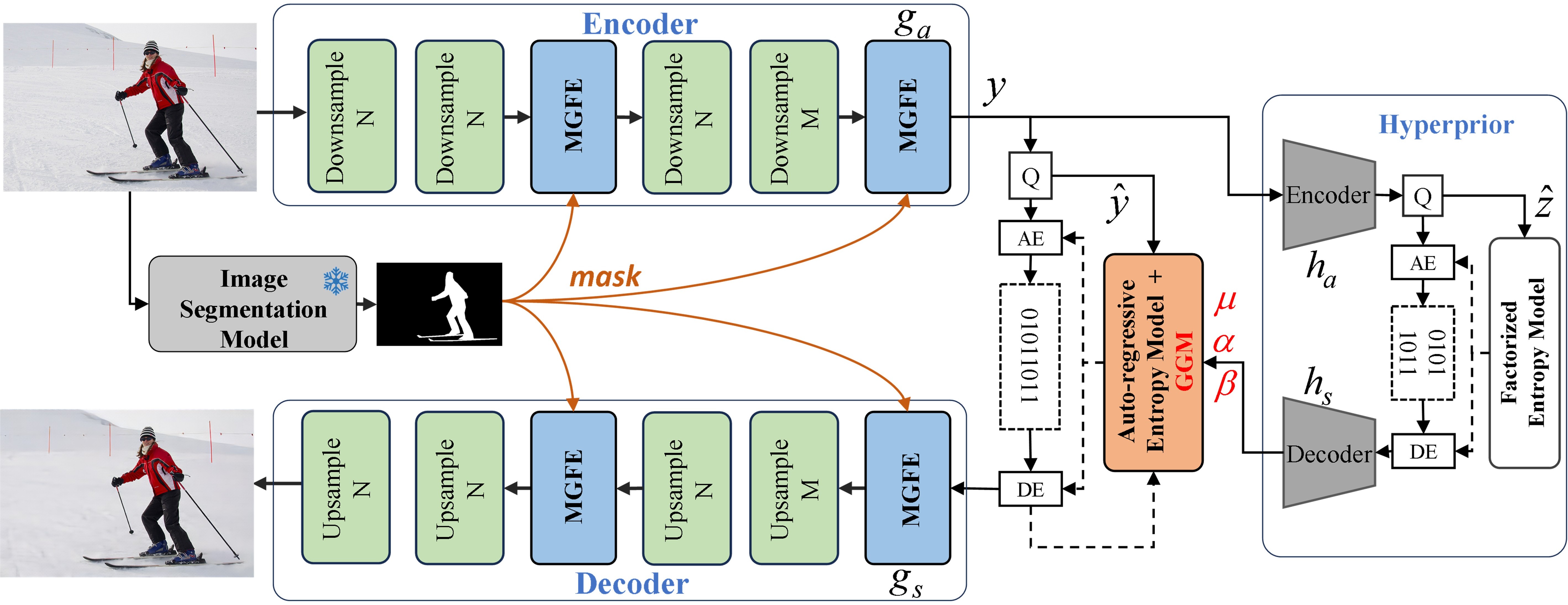} 
			\caption{Our overall framework. MGFE represents Mask-guided Feature Enhancement module as in~\cite{hu2025}, which is used for performing implicit bit allocation and region-adaptive feature enhancement. AE and AD are arithmetic en/de-coder.}
			\label{fig:network}
			\vspace{-0.15cm}
		\end{figure}
\begin{figure}[htbp]
			\centering
			\includegraphics[scale=0.41]{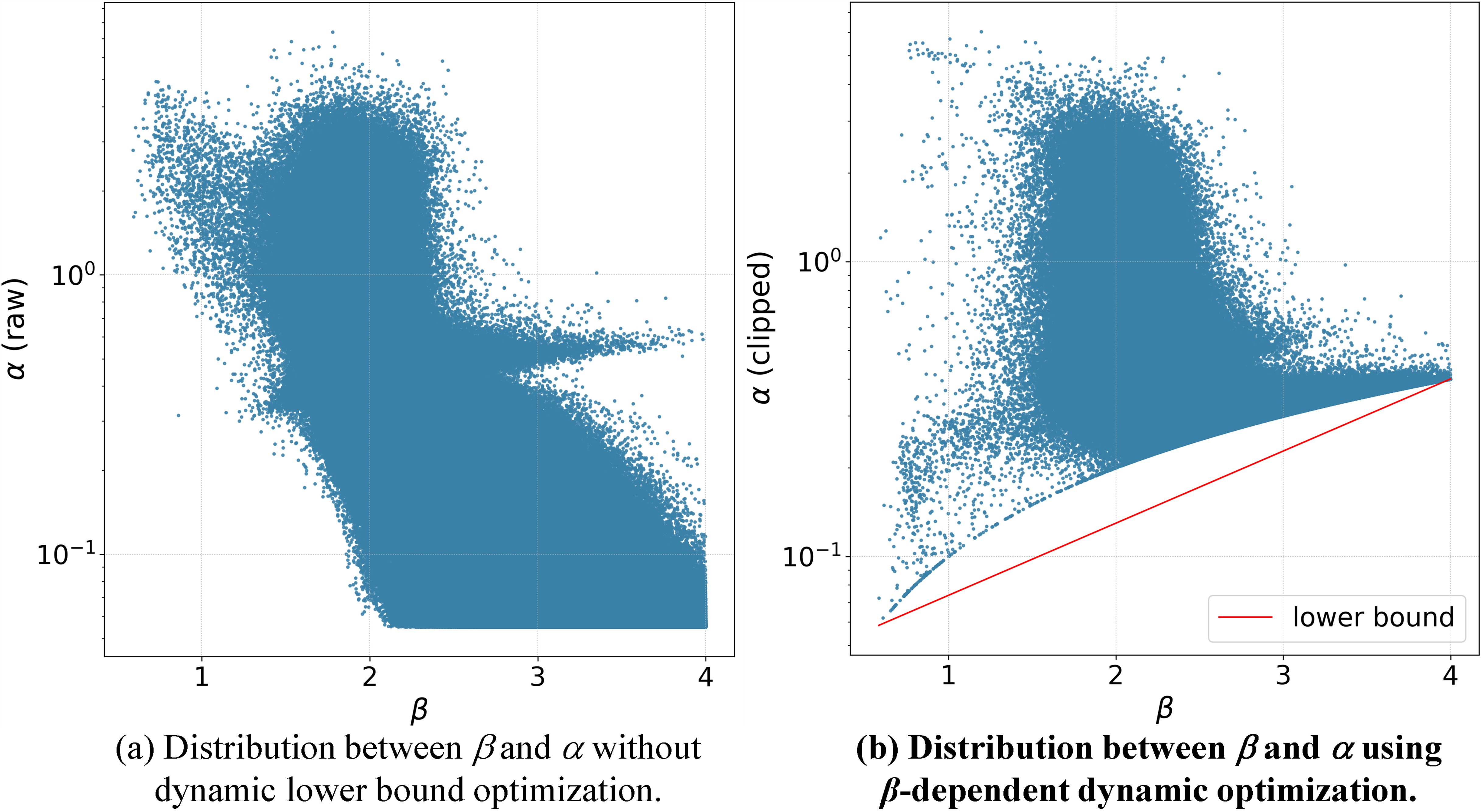} 
			\caption{Illustration of the dynamic lower bound for scale $\mathbf{\alpha}$. These distributions are mainly derived from 100 images and a set of $1\times10^{7}$ samples in the Pascal VOC2012 dataset.}
			\label{fig:alphabetavisual}
		\end{figure}

		\textbf{Dynamic lower bound for scale $\mathbf{\alpha}$:} 
		A train-test mismatch arises from the discrepancy between additive uniform noise during training and hard rounding during testing. While this mismatch in the distortion term is often addressed, it persists in the rate term. This issue is exacerbated when the scale $\mathbf{\alpha}$ becomes very small, as shown in Figs.~\ref{fig:alphas_r}(a)-(c). During testing, the probability mass over the interval $[-0.5, 0.5]$ approaches 1.0, misleading the model into estimating that encoding the symbol ‘0’ requires almost no bits. During training, however, the model needs to estimate the bitrate for the entire continuous interval, resulting in a systematic overestimation of the rate. Additionally, the latent distribution exhibits strong spatial heterogeneity in ROI-based compression. This structural divergence introduces a unique optimization challenge beyond the general train-test mismatch addressed by prior bounds (e.g., fixed bounds for Gaussian models~\cite{he2022elic,zou2022devil,zhang2023uniform} or the theoretically-derived $\mathbf{\beta}$-dependent bound for GGM~\cite{zhang2025generalized}.
		Specifically, without a suitable constraint tailored to this heterogeneity, the joint optimization of $\mathbf{\alpha}$ and $\mathbf{\beta}$ becomes unstable. As visualized in Fig.~\ref{fig:alphabetavisual}(a), numerous small $\mathbf{\alpha}$ values erroneously cluster in regions of relatively large $\mathbf{\beta}$. This ill-posed parameter coupling not only exacerbates rate estimation errors but, more critically, degrades the model's capacity to accurately represent the distinct statistics of foreground and background regions, thereby limiting overall coding performance.
		
		To tackle this instability while maintaining adaptation to the distribution shape, we introduce a dynamic lower bound for $\mathbf{\alpha}$ that explicitly links it to $\mathbf{\beta}$ through a simple yet effective linear constraint:
		\begin{equation}
			\mathbf{\alpha}_{min} = \max \{ \mathbf{\alpha}, \, \zeta \cdot \mathbf{\beta} \},
		\end{equation}
		where $\zeta$ is a hyperparameter. This design ensures that the minimum allowed value of $\mathbf{\alpha}$ is no longer a fixed constant but adapts to the distribution shape indicated by $\mathbf{\beta}$. Given that the latent space distributions in ROI-based coding are typically sharp-peaked and heavy-tailed, a smaller $\mathbf{\beta}$ (corresponding to a heavier-tailed distribution) permits a correspondingly lower $\mathbf{\alpha}$ to accurately model values clustered around zero. Conversely, a larger $\mathbf{\beta}$ necessitates a higher $\mathbf{\alpha}$, preventing the GGM from generating unrealistically narrow distributions that exacerbate train-test mismatch. As shown in Fig.~\ref{fig:alphabetavisual}(b), when this dynamic constraint is applied, $\mathbf{\alpha}$ values are positively correlated with $\mathbf{\beta}$, and excessively small $\mathbf{\alpha}$ values are effectively suppressed. This mechanism reduces the train-test discrepancy, enhancing the overall coding performance. 
		
		\subsection{Overview of Network Framework}
		Based on the RDO paradigm mentioned in Eq. \eqref{eq7}, we develop an efficient ROI-based image compression model, as illustrated in Fig.~\ref{fig:network}. It adopts the model~\cite{hu2025} as network backbone but utilizes only a single decoder to simultaneously reconstruct both the foreground and background of the image, thereby reducing model complexity. An original image $\mathbf{x} \in{\mathbb{R}}^{H \times W \times 3}$ is first fed into the encoder network $g_a$ and pretrained segmentation models. Masks can be generated by the pre-trained segmentation model, such as semantic segmentation~\cite{wang2020solov2}. To leverage the generated mask $\mathbf{M}$ for implicit bit allocation, MGFE modules are adopted (a detailed explanation is provided in Appendix~\ref{appsub3}). It precisely distinguishes foreground from background regions, enabling high-quality region-adaptive coding. In order to more accurately capture latent space distributions in ROI-based compression, we introduce the GGM to replace the GM. 
		
		Latent features $\mathbf{y}$ represented by the encoder network $g_a$ are then further transformed into hyperprior latent features $\mathbf{z}$ through the hyper-prior encoder $h_a$. The features $\mathbf{y}$ and $\mathbf{z}$ are input into the quantizer $Q$ to yield results $\hat{\mathbf{y}}$ and $\hat{\mathbf{z}}$. A simple factorized entropy model is used to encode the quantized $\hat{\mathbf{z}}$. Subsequently, $\hat{\mathbf{z}}$ is fed into the hyperprior decoder $h_s$ after decoding to derive  three  parameters, such as  mean $\mathbf{\mu}$,  scale  $\mathbf{\alpha}$, and  shape  $\mathbf{\beta}$. To facilitate a GGM for entropy encoding, we use zero-center quantization $\lceil \mathbf{y} - \mathbf{\mu} \rfloor + \mathbf{\mu}$. Owing to the implicit bit allocation strategy,  quantized latent features $\hat{\mathbf{y}}$ retain intact background information. Finally, we deploy a decoder $g_s$ to reconstruct high-quality images $\hat{\mathbf{x}}$.

		\subsection{Training Methods}
		\textbf{Differentiable  optimization of $P (a, b)$ function:}
		When optimizing the parameters $\mathbf{\mu}$ $\mathbf{\alpha}$, $\mathbf{\beta}$ through gradient descent, the derivative of $c_\beta(\mathbf{y})$ can be calculated through
		\begin{equation}
			\begin{aligned}
				\frac{\partial c_\beta(\mathbf{y})}{\partial \mathbf{y}} &= f_\beta(\mathbf{y}),\\
				\frac{\partial c_\beta(\mathbf{y})}{\partial\mathbf{\beta}} &= \frac{\operatorname{sgn}(\mathbf{y})}{2}\left(-\frac{1}{\mathbf{\beta}^2}P' + \frac{|\mathbf{y}|\ln|\mathbf{y}|}{\Gamma(1/\mathbf{\beta})}e^{-|\mathbf{y}|^\mathbf{\beta}}\right),
			\end{aligned}
		\end{equation}
		According to Eq. \eqref{eq5}, we can obtain
		\begin{equation}
			\begin{aligned}
				P^{'} = \frac{\partial P}{\partial a}\,\frac{da}{d\mathbf{\beta}} + \frac{\partial P}{\partial b}\,\frac{db}{d\mathbf{\beta}}, 
			\end{aligned}
		\end{equation}
		where two main derivative solutions are $\frac{\partial P}{\partial b} = \frac{b^{\,a-1} e^{-b}}{\Gamma_a}$ and $\frac{\partial P(a, b)}{\partial a} = \frac{\partial(\frac{\gamma(a,b)}{\Gamma(a)})}{\partial a}$, respectively. By applying the fractional differentiation rule, $\frac{\partial P(a,b)}{\partial a}$  is decomposed into
		\begin{equation}
			\begin{aligned}
				\frac{\partial P(a,b)}{\partial a} = \frac{\dfrac{\partial \gamma(a,b)}{\partial a} \,\Gamma(a)\,-\,\gamma(a,b)\,\dfrac{d \Gamma(a)}{d a}}{\Gamma(a)^2},
			\end{aligned}
		\end{equation}
		To compute this expression, we require the derivatives $\dfrac{\partial \gamma(a,b)}{\partial a}$ and $\dfrac{d \Gamma(a)}{d a}$. The derivative of the Gamma function is $\frac{d \Gamma(a)}{d a} = \Gamma(a)\psi(a)$, where $\psi(a)$ is the digamma function. For the gamma function $\gamma(a,b)$, $\frac{\partial \gamma(a,b)}{\partial a} = \int_0^b t^{a-1} e^{-t} \ln(t) dt$. This integral lacks a closed-form solution and poses challenges for stable and efficient numerical computation~\cite{blahak2010efficient}. Consequently, within automatic differentiation frameworks like PyTorch, the function $P(a,b)$ is non-differentiable with respect to the parameter $a$.
		
		To enable gradient-based stable optimization, we circumvent direct numerical integration. Instead, we propose an efficient central difference scheme to approximate the gradient of $\gamma(a, b)$, thereby restoring differentiability during training. 
		\begin{equation}
			\begin{aligned}
				\frac{\partial}{\partial a}\gamma(a,b) \approx \frac{\gamma(a+\epsilon,b) - \gamma(a-\epsilon,b)}{2\epsilon},
			\end{aligned}
			\label{eqgamma}
		\end{equation}
		The step size is $\epsilon = 1 \times 10^{-5}$. Specifically, our finite difference method has a constant complexity of $\mathcal{O}(1)$, while numerical integration typically requires $\mathcal{O}(N)$ operations, where $N$ is the number of integration steps (a detailed error bound analysis is provided in Appendix~\ref{appsub2}). The gradient computation via finite differences for the GGM is detailed in Algorithm~\ref{alg:ggd_cdf}.

		\definecolor{lightgray}{rgb}{0.7,0.7,0.7}
		\algrenewcommand{\algorithmiccomment}[1]{\textcolor{lightgray}{// #1}}
		
		\begin{algorithm}[t]
			\footnotesize
			\caption{Gradients computation via finite differences for the GGM}
			\label{alg:ggd_cdf}
			\begin{algorithmic}[1]
				\Require Input latent tensors $\mathbf{x}$; shape parameter $\mathbf{\beta}$; valid interval ($\mathbf{\beta}_{\min}$, $\mathbf{\beta}_{\max}$); minimum value $\varepsilon$; step size $\epsilon_{\mathrm{fd}}$  
				
				\Function{Forward}{$\mathbf{x},\mathbf{\beta}$} \Comment{Compute CDF}
				\State \Comment{Ensure it is positive and within an effective range}
				\State $\mathbf{\beta} \gets \operatorname{clamp}\big(\operatorname{softplus} (\mathbf{\beta}),\;\mathbf{\beta}_{\min},\;\mathbf{\beta}_{\max}\big)$ 
				\State $x_{\mathrm{abs}} \gets \max(|x|,\varepsilon)$
				\State $a \gets 1/\mathbf{\beta}$; $b \gets x_{\mathrm{abs}}^{\,\mathbf{\beta}}$ \Comment{Parameters transform}
				\State $\mathrm{cdf} \gets \tfrac{1}{2}\big(1 + \operatorname{sign}(x) \cdot P(a,b)\big)$  \Comment{Output CDF} 
				\State \textbf{save} $(x_{\mathrm{abs}}, \mathbf{\beta}, a, b)$ \Comment{For backward}
				\State \Return $\mathrm{cdf}$
				\EndFunction
				
				\Function{Backward}{$\mathrm{grad\_out}$} \Comment{Compute $x, \mathbf{\beta}$ gradients}
				\State \textbf{load} $(\mathbf{x}_{\mathrm{abs}}, \mathbf{\beta}, a, b)$
				\State $\Gamma_a \gets \exp(\lgamma(a))$ \Comment{Gamma value}
				\State $\displaystyle \frac{\partial P}{\partial b} \gets \frac{b^{\,a-1} e^{-b}}{\Gamma_a}$; $\displaystyle \frac{db}{dx} \gets \mathbf{\beta}\, \mathbf{x}_{\mathrm{abs}}^{\,\mathbf{\beta}-1}$
				\State \Comment{Output $\mathbf{x}$ gradients}
				\State $\displaystyle \mathrm{grad\_\mathbf{x}} \gets \mathrm{grad\_out} \cdot  \frac{\operatorname{sign}(\mathbf{x})} 2
				\big(\frac{\partial P}{\partial b} \cdot \frac{db}{dx}\big)$ 
				
				\State $v_{+} \gets P\big(a+\epsilon_{\mathrm{fd}},\,b\big)\cdot\exp\big(\lgamma(a+\epsilon_{\mathrm{fd}})\big)$
				\State $v_{-} \gets P\big(a-\epsilon_{\mathrm{fd}},\,b\big)\cdot\exp\big(\lgamma(a-\epsilon_{\mathrm{fd}})\big)$
				\State \Comment{Approximate via central finite differences}
				\State $\displaystyle \frac{\partial \gamma(a,b)}{\partial a} \approx \frac{v_{+}-v_{-}}{2\epsilon_{\mathrm{fd}}}$ 
				\State $\displaystyle \frac{\partial P}{\partial a} \gets
				\frac{1}{\Gamma_a} \frac{\partial \gamma(a,b)}{\partial a} - P(a,b)\,\psi(a)$
				
				\State $\displaystyle \frac{da}{d\beta} \gets -\mathbf{\beta}^{-2}$; $\displaystyle \frac{db}{d\beta} \gets b\cdot\log(\mathbf{x}_{\mathrm{abs}})$
				\State $\displaystyle \frac{\partial P}{\partial \mathbf{\beta}} \gets
				\frac{\partial P}{\partial a}\,\frac{da}{d\beta} + \frac{\partial P}{\partial b}\,\frac{db}{d\beta}$
				\State $\displaystyle \mathrm{grad\_\beta} \gets \mathrm{grad\_out}\cdot \frac{\operatorname{sign}(\mathbf{x})} 2\cdot
				\frac{\partial P}{\partial \mathbf{\beta}}$ \Comment{Output $\mathbf{\beta}$ gradients}
				\State \Return $(\mathrm{grad\_\mathbf{x}},\ \mathrm{grad\_\mathbf{\beta}})$
				\EndFunction
			\end{algorithmic}
		\end{algorithm}

		\textbf{Training Loss:} 
		ROI-based neural image compression is fundamentally an RDO problem. Its objective is to minimize a weighted sum of the rate and a distortion term that is evaluated separately for the foreground and background. Building on Eq. \eqref{eq7}, the rate term for our task can be expressed as:
		\begin{equation}
			\begin{aligned}
				\boldsymbol{\mathcal{R}} = \mathbb{E}_{p(\mathbf{x})} \mathbb{E}_{q_\phi(\hat{\mathbf{y}}, \hat{\mathbf{z}}|\mathbf{x})} [ -\log p_\theta(\hat{\mathbf{y}}|\hat{\mathbf{z}}) - \log p_\theta(\hat{\mathbf{z}}) ],
			\end{aligned}
		\end{equation}
		where $p_{\hat{\mathbf{y}}}$ and $p_{\hat{\mathbf{z}}}$ denote the probability distributions of the quantized latents $\hat{\mathbf{y}}$ and $\hat{\mathbf{z}}$, respectively. Correspondingly, the distortion term $\mathcal{D}$ is decomposed into a foreground component $\mathcal{D}_{FG}$ and a global component $\mathcal{D}_{G}$. Thus, our overall loss function is defined as
		\begin{equation}
			\begin{aligned}
				\boldsymbol{\mathcal{L}} = (\mathcal{D}_{FG} + k_1 \cdot \mathcal{D}_{G} + \mathcal{L}_{b{\_}{rec}}) + \lambda \cdot \boldsymbol{\mathcal{R}}, 
			\end{aligned}
		\end{equation}
		$\mathcal{L}_{b{\_}{rec}}$ represents a deep perception loss of global images.
		
		To achieve high-fidelity foreground reconstruction while maintaining satisfactory background quality, we employ region-specific loss functions. The distortion terms are defined as
		\begin{equation}
			\begin{aligned}
				& \mathcal{D}_{FG} = \text{MSE}_{ROI}\left(\mathbf{x} \times \mathbf{M}, \hat{\mathbf{x}} \times \mathbf{M} \right), \\
				& \mathcal{D}_{G} = \text{MS-SSIM} \left( \mathbf{x}, \hat{\mathbf{x}} \right), \\
			\end{aligned}
		\end{equation}
		Here, $\text{MSE}_{ROI}$ is the mean squared error loss within the ROI. To align with human visual perception and preserve detail, the global loss $\mathcal{D}_{G}$ integrates a multi-scale structural similarity (MS-SSIM) loss~\cite{1292216}. Additionally, we incorporate perceptual losses $\boldsymbol{\mathcal{L}}_{per}$ and style losses $\boldsymbol{\mathcal{L}}_{sty}$ following~\cite{gfpgan} to enhance the naturalness and consistency of output images:
		\begin{equation}
			\begin{aligned}
				\boldsymbol{\mathcal{L}}_{b{\_}{rec}} = k_2 \cdot \boldsymbol{\mathcal{L}}_{per} + k_3 \cdot \boldsymbol{\mathcal{L}}_{sty}, 
			\end{aligned}
		\end{equation}
		where hyper-parameters are set as $k_1=100$, $k_2=0.02$, and $k_3=50$.
		
		\begin{table*} [!ht]
			\centering
			\renewcommand{\arraystretch}{1.8}
			\vspace{-0.3cm} 
			\caption{Average BD-rate ($\%$) saving, BD-PSNR ($dB$) gain, and BD-mAP ($\%$) gain compared to Zou2022~\cite{zou2022devil} across COCO2017 and HRSOD dataset. Symbol (``-'') indicates better performance than Zou2022 for BD-rate gain, while indicates the opposite for BD-mAP and BD-PSNR gain.}
			\label{tab:mapBD}
			\scalebox{0.85}{
				\begin{tabular}{ccccccccccc}  \hline
					\multicolumn{1}{c}{\multirow{3}{*}{\textbf{Methods}}} & \multicolumn{6}{c}{\textbf{COCO2017 dataset}} & \multicolumn{4}{c}{\textbf{HRSOD dataset}} \\ 
					\cmidrule(r){2-7} \cmidrule(l){8-11}
					&  \multicolumn{2}{c}{\textbf{ROI-PSNR}} & \multicolumn{2}{c}{\textbf{Segmentation Accuracy}} & \multicolumn{2}{c}{\textbf{Detection Accuracy}} & \multicolumn{2}{c}{\textbf{PSNR}}  & \multicolumn{2}{c}{\textbf{ROI-PSNR}} \\ 
					& BD-rate $\downarrow$ & BD-PSNR $\uparrow$  & BD-rate $\downarrow$ & BD-mAP $\uparrow$  & BD-rate $\downarrow$ & BD-mAP $\uparrow$ & BD-rate $\downarrow$ & BD-PSNR $\uparrow$ & BD-rate $\downarrow$ & BD-PSNR $\uparrow$ \\ \hline
					Zou2022~\cite{zou2022devil}  & -- & --  & -- & -- & -- & -- & -- & -- & -- & --  \\ 
					Ballé2018~\cite{balle2018variational} & 136.02 & -3.416 & 75.44 & -3.215 & 76.76 & -2.995 & 150.63 & -4.488 & 152.08 & -4.608 \\ 
					BPG(YUV444)  & 44.16 & -1.343 & 78.30 & -3.215 & 75.44 & -3.674 & 38.17 & -1.209 & 40.57 & -1.325 \\ 
					Cheng2020~\cite{cheng2020learned} & 19.32 & -0.666 & -3.29 & 0.042 & -2.55 & -0.049 & -1.65 & 0.054 & -0.18 & -0.001 \\ 
					VVC~\cite{vtm}  & 5.95 & -0.261 & 26.25 & -1.199 & 24.08 & -1.330 & 13.11 & -0.464 & 15.05 & -0.504 \\
					Liu2023~\cite{liu2023learned}  & -6.55 & 0.260  & 5.55 & -0.241 & -0.57 & -0.261 & -10.40 & 0.422 & -9.49 & 0.399 \\
					Liu2024~\cite{liu2024region} & -10.52 & 0.403  & -14.84 & 0.636 & -14.61 & 0.699 & -6.04 & 0.231 & -4.92 & 0.195 \\ 
					Hu2025~\cite{hu2025} &  -81.98 & 5.897 & -89.63 & 5.032 & -87.49 & 5.725 & -10.14 & 0.309 & -86.94 & 7.531 \\ \hline 
					\textbf{Our} & \textbf{-85.00} & \textbf{6.294} & \textbf{-99.78} & \textbf{5.443} & \textbf{-99.99} & \textbf{6.298} & \textbf{-28.46} & \textbf{0.949} & \textbf{-89.89} & \textbf{7.900} \\ \hline
			\end{tabular}}
		\end{table*}
		\section{Experiments}
		\label{5}
		\subsection{Experiment Details}
		\textit{\textbf{Experimental Data}}. We conduct experiments using the COCO2017 dataset~\cite{lin2014microsoft}, which provides training, validation, and test splits. From the training set of approximately 118,000 images, we randomly select 15,000 images, cropping each to $256\times256$ pixels for training. Corresponding foreground masks are generated by binarizing the instance segmentation annotations at the pixel level. We select mask-image pairs where the foreground occupies 8\% to 80\% of the image area for both training and testing, resulting in 3,443 pairs from the 5,000-image validation set for evaluation. 
		
		To assess generalization, we additionally employ the HRSOD dataset~\cite{zeng2019towards}, a benchmark for high-resolution salient object detection containing 8,398 training and 2,100 testing images (resolutions between 1200 and 2560 pixels). Following a similar selection procedure, we randomly sample 115 mask-image pairs within the same foreground ratio range of $[0.08, 0.8]$.
		
		\textit{\textbf{Experimental Settings.}} In our experiments, the quantization strategy differs by model: GMM and GLLMM employ non-zero-center quantization ($\lfloor y \rceil$), while all other probability models, including our GGM, use zero-center quantization ($\lfloor \mathbf{y} - \mathbf{\mu} \rceil + \mathbf{\mu}$). All models are optimized using Adam with an MSE loss, a batch size of 8, and are trained for 450 epochs. We train multiple models across rate points by varying the Lagrange multiplier $\lambda \in \{32, 64, 128, 512 \}$. The learning rate is initially $1\times10^{-4}$ and decays to $1\times10^{-5}$ for the final 100 epochs. The channel counts $N$ and $M$ in the compression network are set to 192 and 320, respectively. The hyperparameter $\zeta$ is set to 0.1. Our implementation is based on the CompressAI PyTorch library~\cite{begaint2020compressai}, and experiments are conducted on a single NVIDIA V100 GPU.
		
		\begin{figure*}[!t] 
			\centering    
			\includegraphics[scale=0.22]{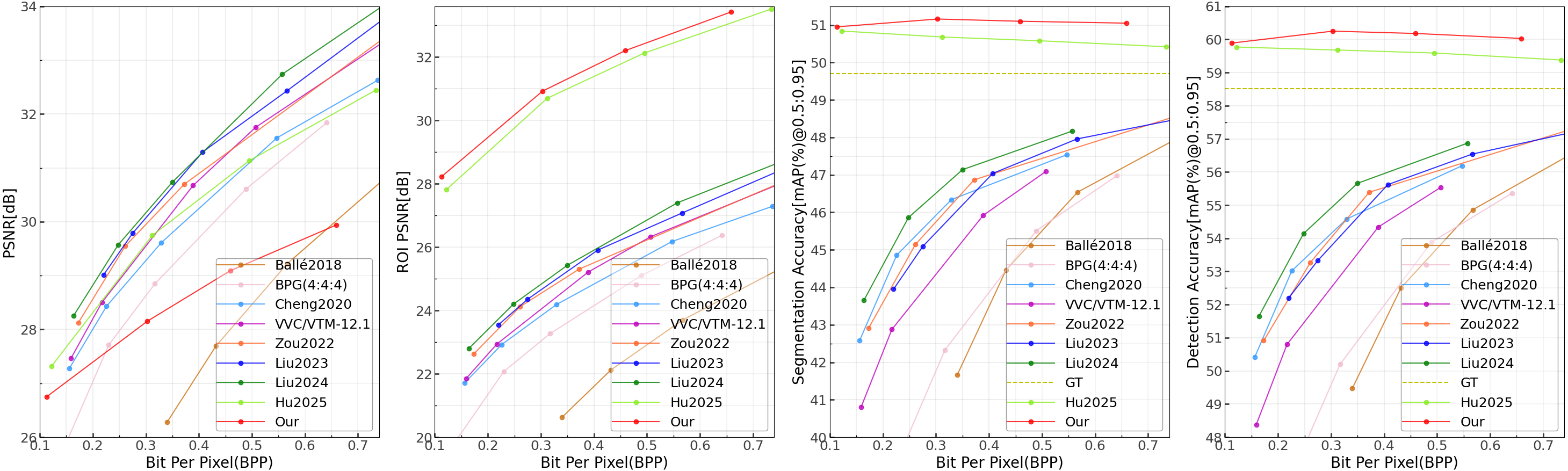} 
			\caption{RD performance averaged in terms of PSNR, ROI-PSNR, and mAP@0.50:0.95 on COCO2017 dataset.}
			\label{fig:psnrdetseg}
			\vspace{-0.1cm}
		\end{figure*}
		Our performance benchmarks include NIC methods (Ballé2018~\cite{balle2018variational}, Cheng2020~\cite{cheng2020learned}, Zou2022~\cite{zou2022devil}, Liu2023~\cite{liu2023learned}, Liu2024~\cite{liu2024region}), classic codecs (BPG~\cite{bpg}, VVC/VTM-12.1~\cite{vtm}), and the recent ROI method Hu2025~\cite{hu2025}. Reconstruction quality is evaluated using RD curves, plotting bit-per-pixel (BPP) against both overall PSNR and ROI-PSNR. Performance is quantitatively compared using BD-rate and BD-PSNR metrics based on the Bjøntegaard-Delta method~\cite{Bjontegaard}. Beyond pixel fidelity, we assess the utility of compressed images for machine vision (MV) by evaluating object detection and instance segmentation performance. The mean Average Precision (mAP) with an IoU threshold of 0.50:0.95 serves as the primary MV metric. We plot BPP versus mAP and employ analogous BD-mAP and BD-rate metrics for quantitative comparison.
		
		\begin{figure*}[!t] 
			\centering    
			\includegraphics[scale=0.48]{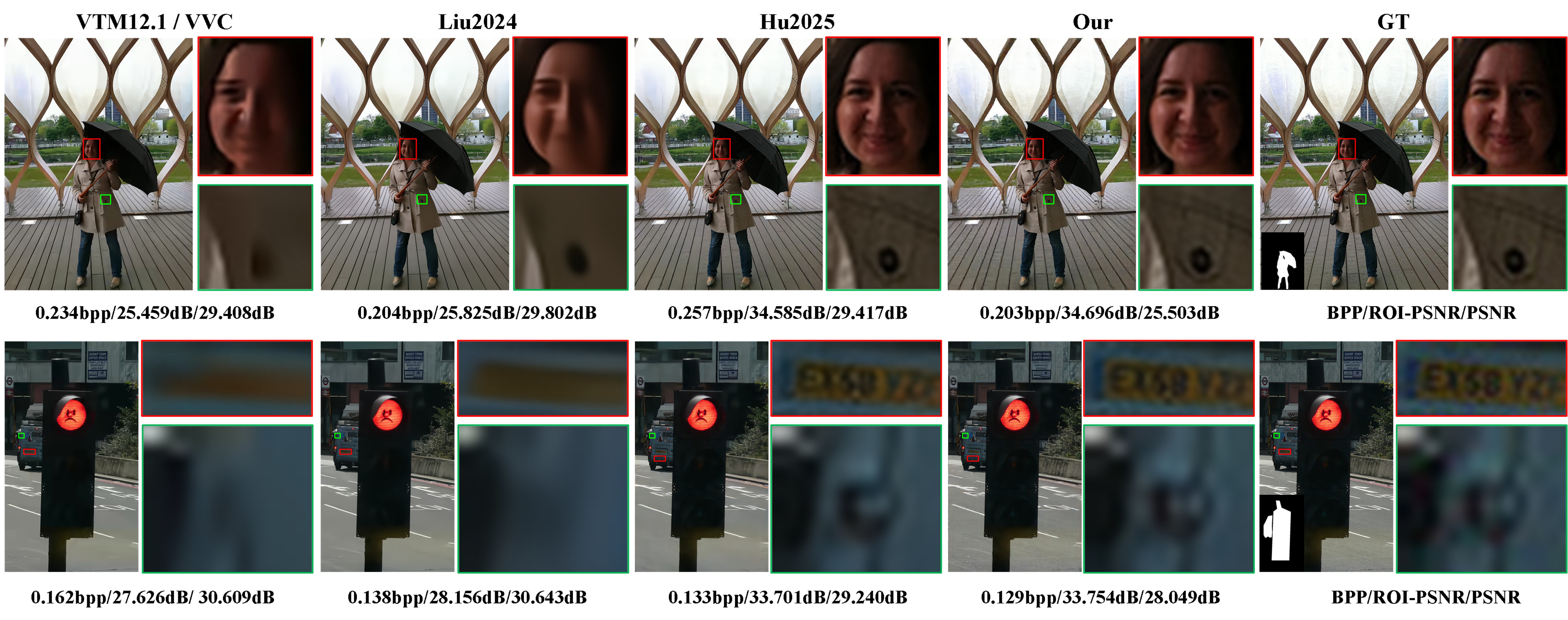} 
			\caption{Visual comparison of compression results using different methods on the COCO2017 dataset. We also place a mask on the GT image to visually compare the reconstruction effect of important areas.}
			\label{fig:psnrvisual}
			\vspace{-0.1cm}
		\end{figure*}
		\subsection{Compression Performance} 
		To validate the compression performance, we plot RD curves for various methods. As shown in Fig.~\ref{fig:psnrdetseg}, our method achieves the best RD performance in the ROI, outperforming the recent state-of-the-art method (Hu2025~\cite{hu2025})  and conventional NIC methods. This advantage can be attributed to our elaborate GGM and implicit bit allocation method. It is noted that, due to the dedicated focus on ROI fidelity within a single-decoder framework, our method’s performance on the entire image is comparatively lower. In contrast, Hu2025 employs a dual-decoder architecture, which achieves a more balanced reconstruction between foreground and background. Fig.~\ref{fig:psnrvisual} provides a visual comparison. The images reconstructed by our method—optimized with a combined perceptual and MS-SSIM loss—retain high visual quality. In particular, reconstruction within the ROI is notably superior to other methods and closest to the original uncompressed image (GT). While the Hu2025 method also demonstrates good fidelity, its recovery of high-frequency textures and color accuracy remains less precise.
		
		For a quantitative assessment, we employ BD-rate and BD-PSNR metrics, using Zou2022~\cite{zou2022devil} as the baseline (Tab.~\ref{tab:mapBD}). Our method achieves a BD-rate reduction of 85\% and a BD-PSNR gain of 6.294 dB, substantially surpassing all compared methods. Hu2025 shows a 2.25\% lower BD-rate reduction and a 0.311 dB smaller BD-PSNR gain relative to our results, further confirming the effectiveness of our approach for ROI coding. The foundational method Balle2018~\cite{balle2018variational} exhibits a BD-rate increase of 136.02\% and a BD-PSNR drop of 3.416 dB under this evaluation.
		We further evaluate generalization on the high-resolution HRSOD dataset~\cite{zeng2019towards} (Tab.~\ref{tab:mapBD}). Our GGM method again achieves optimal performance, with BD-rate reductions of 28.46\% (full images) and 89.89\% (target region) compared to the Zou2022 method~\cite{zou2022devil}. This is primarily due to the distribution characteristics of the HRSOD dataset, which contains relatively few high-frequency textures in the background regions aside from the foreground objects. These results consistently demonstrate the superiority and general applicability of our GGM for ROI-based compression.
		
		\begin{figure*}[!t] 
			\centering    
			\includegraphics[scale=0.55]{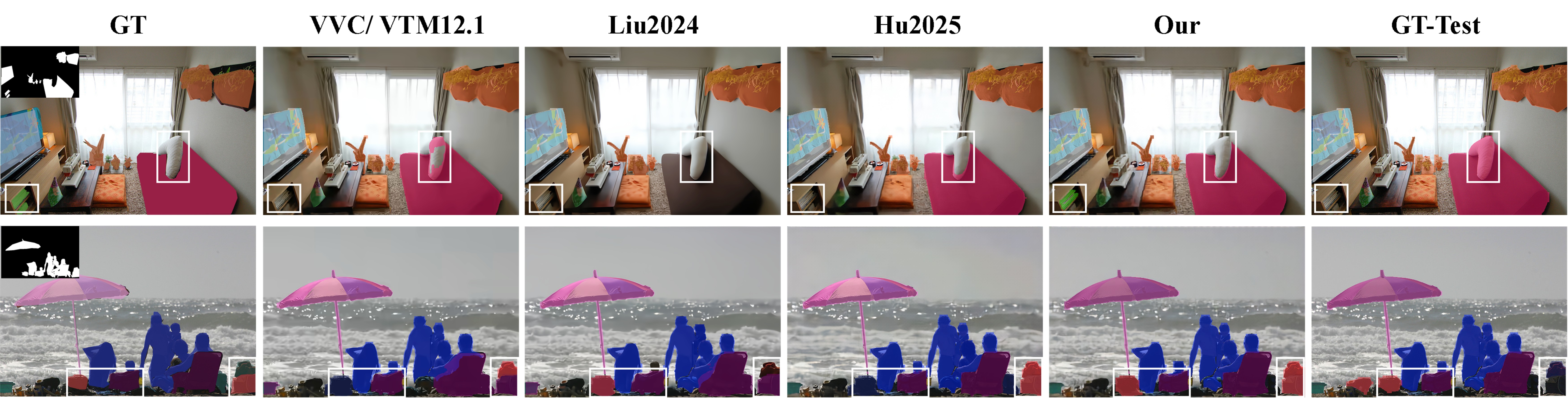} 
			\caption{Segmentation results on COCO2017 for different compression methods. White boxes indicate major errors (missegmentation/missed detection). GT-test represents segmentation results of original images after testing. The GT image includes its annotation mask for reference.}
			\label{fig:segvisual}
			\vspace{-0.1cm}
		\end{figure*}
		\subsection{Machine-vision Performance}
		The primary objective of this study is to enhance the utility of compressed images for MV tasks. We therefore evaluate the impact of different compression methods on two representative MV tasks: object detection and instance segmentation. Given the strong performance and available pre-trained variants of YOLOv11~\cite{khanam2024yolov11}, we adopt it as our MV task network. To visually compare performance, we plot the curves of recognition accuracy and BPP for both tasks, as shown in Fig.~\ref{fig:psnrdetseg}. Our reconstructed images achieve the highest mAP across bitrates, outperforming both the recent ROI-based method Hu2025~\cite{hu2025} and even the GT-test images. This advantage stems from our method’s high-fidelity reconstruction in ROIs and its suppression of distracting background textures, which collectively reduce misidentification and missed detection. Fig.~\ref{fig:segvisual} visualizes segmentation results of different compression methods. Our method produces the most accurate segmentation masks, exhibiting fewer misclassifications and missed detections while maintaining high accuracy across objects of varying sizes. Although the Hu2025 method and the GT-test also achieve reasonable performance, they show more noticeable segmentation errors and instances of missed detection.
		
		For a quantitative comparison, we compute BD-rate and BD-mAP metrics using Zou2022~\cite{zou2022devil} as the baseline (Table~\ref{tab:mapBD}). For segmentation tasks, our method achieves a BD-rate reduction of 99.78\% and a BD-mAP gain of 5.443\%, significantly surpassing all other methods. Notably, it exceeds the Hu2025 method by 10.15\% in BD-rate reduction and 0.411\% in BD-mAP gain. Similarly, for object detection, our method attains a BD-rate reduction of 99.99\% and a BD-mAP gain of 6.298\%, outperforming Hu2025 by 12.5\% and 0.573\%, respectively. These results consistently demonstrate that our method best preserves the information critical for downstream MV tasks.
		
		\begin{figure*}[!t] 
			\centering    
			\includegraphics[scale=0.55]{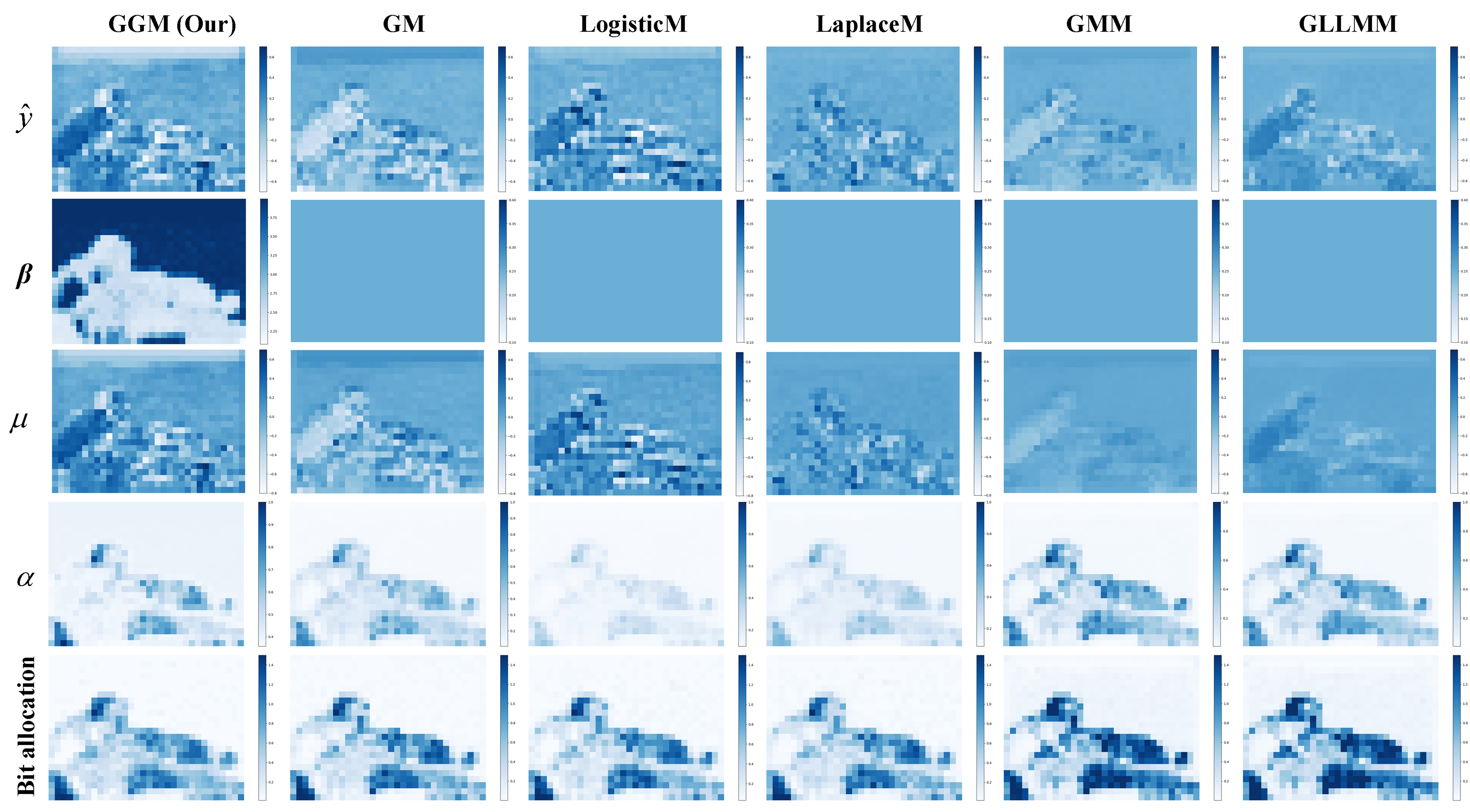} 
			\caption{Visualization of shape parameters $\mathbf{\beta}$, mean $\mathbf{\mu}$, scale parameters $\mathbf{\alpha}$, quantized latent features $\hat{y}$, and bit allocation maps for different PDMs across Pascal VOC2012 dataset. }
			\label{fig:diffdm}
			\vspace{-0.1cm}
		\end{figure*}
		\subsection{Comparison of Different Probability Distribution Models}
		To analyze the impact of different probability distribution models (PDMs) on coding performance, we conduct a comparative experiment by replacing the GGM in our compression network with several classic PDMs: the GM, the Laplacian Model (LaplaceM), the Logistic Model (LogisticM), the GMM, and the GLLMM. Performance is evaluated from three aspects: PSNR, ROI-PSNR, and model complexity. We adopt the GM as the baseline to calculate and compare their compression performance in metrics such as BD-rate, BD-PSNR, parameter counts, and Multiply–Accumulate Operations (MACs). The experimental results (Table~\ref{tab:mapPM}) demonstrate that our GGM, with its learned shape parameter $\mathbf{\beta}$, accurately models latent distributions and achieves the best coding performance. Compared to the GM baseline, our GGM achieves a BD-rate reduction of 41.53\% and a BD-PSNR gain of 1.131 dB for the overall image, and a BD-rate reduction of 9.58\% with a BD-PSNR gain of 0.303 dB for the ROI. The LaplaceM and LogisticM show slightly weaker performance than the GM, indicating a poorer fit to the actual latent distributions. GMM and GLLMM are designed to model complex distributions via mixture components. However, despite their theoretical approximation capacity, they suffer from optimization difficulties in our specific task. Our latent features primarily follow a sharp-peaked and heavy-tailed distribution, causing instability in parameter estimation for mixture models during end-to-end training. This fails to converge to the optimal distribution effectively. In detail, GMM and GLLMM increase BD-rate by 419.83\% and 236.90\%, and decrease BD-PSNR by 8.323 dB and 7.772 dB, respectively. In terms of complexity, our network with GM has 40.599M parameters and 173.712G MACs/pixel. The GLLMM exhibits the highest complexity, adding 6.395M parameters and 0.576G MACs/pixel, substantially exceeding our GGM-based model.
		
		\begin{table} [!ht]
			\centering
			\renewcommand{\arraystretch}{2.5}
			\small 
			\caption{Average performance gain of different PDMs compared to the GM across COCO2017 dataset.}
			\label{tab:mapPM}
			\scalebox{0.58}{
				\begin{tabular}{c@{\hspace{0.5em}}cccccc}  \hline
					\multicolumn{1}{c}{\textbf{PDMs}} & \multicolumn{2}{c}{\textbf{PSNR}} & \multicolumn{2}{c}{\textbf{ROI-PSNR}} & \multicolumn{2}{c}{\textbf{Model Complexity}} \\ 
					\cmidrule(r){2-3} \cmidrule(r){4-5}\cmidrule(l){6-7}
					& BD-rate $\downarrow$ & BD-PSNR $\uparrow$ & BD-rate $\downarrow$ & BD-PSNR $\uparrow$  & $\bigtriangleup$ Param(MB) & $\bigtriangleup$ MACs/pixel(G) \\ \hline
					GM  & -- & --  & -- & -- & 40.599 & 173.712  \\ 
					LaplaceM & 8.50 & -0.194 & 3.93 & -0.123 &  0.000  & 0.000    \\ 
					LogisticM  & -5.37 & 0.129 & 8.76 & -0.269 &  0.000  & 0.000      \\ 
					GMM & 419.83 & -8.323 & 139.12 & -2.899 & 1.843  &  0.472     \\ 
					GLLMM  & 236.90 & -7.772 & 119.77 & -3.076 & 12.698 & 3.253  \\ \hline
					\textbf{GGM(Our)} & \textbf{-41.53} & \textbf{1.131} & \textbf{-9.58} & \textbf{0.303} & \textbf{5.395} & \textbf{0.576}  \\ \hline
			\end{tabular}}
		\end{table}
		We further visualize feature maps of different PDMs, quantized latent features $\hat{y}$, and bit allocation maps (Fig.~\ref{fig:diffdm}). Latent features from our GGM retain richer information, contributing to superior reconstruction. The learned $\mathbf{\beta}$ parameter enables accurate distribution modeling. In the bit allocation maps (color intensity indicates bit allocation: deeper = more bits), our GGM-based map shows overall lighter tones, consumes the fewest bits, and allocates zero or minimal bits to many non-important regions. In conclusion, the GGM is particularly well-suited for ROI-based image compression. It accurately models the characteristic latent distribution while maintaining low complexity, thereby enhancing the overall compression performance of the coding network.

		\begin{figure*}[!t] 
			\centering    
			\includegraphics[scale=0.22]{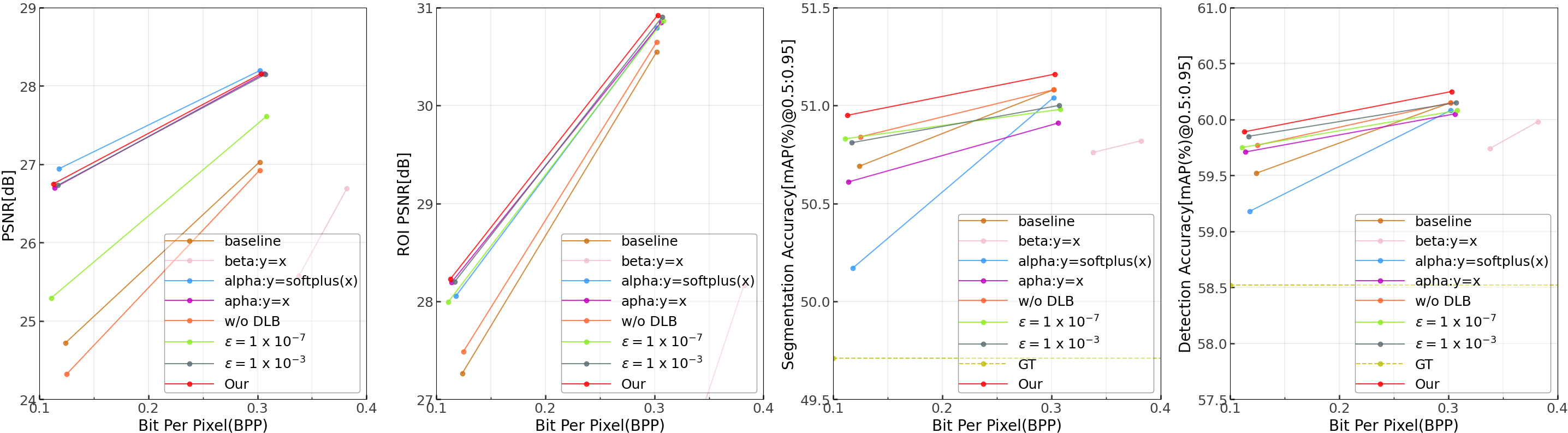} 
			\caption{RD performance of different variants in terms of PSNR, ROI-PSNR, and mAP@0.50:0.95 on COCO2017 dataset.}
			\label{fig:ablatdetseg}
		\end{figure*}
		\subsection{Ablation Study}
		We conduct ablation studies on the key components of our method, primarily focusing on the core parameter settings of GGM. For the shape  $\mathbf{\beta}$, we replace it with the simple linear function $\textit{y = x}$.
		For the scale  $\mathbf{\alpha}$, we test two alternatives: (a) a linear function $\textit{y = x}$ with a fixed lower bound $x \geq 0.11$ (following the GM strategy in~\cite{zhang2023uniform}), and (b) the $\textit{y = softplus(x)}$ function. We also independently experiment with removing the dynamic lower bound (DLB) for the scale $\mathbf{\alpha}$. We then establish a comprehensive baseline where the GGM uses $\textit{y = x}$ for both $\mathbf{\beta}$ and $\mathbf{\alpha}$, and the DLB is removed.
		Additionally, we vary the relative step size $\epsilon$ used in the central difference approximation of $P(a, b)$ to $1 \times 10^{-3}$ and $1 \times 10^{-7}$ to examine its effect on gradient stability and final performance.
		
		\textit{\textbf{Analysis of setting shape parameter $\mathbf{\beta}$}}. The quantitative impact of the shape parameter setting is shown in Table~\ref{tab:Ablation}. Using the linear function $\textit{y = x}$ for $\mathbf{\beta}$ degrades reconstruction quality, especially in the ROI. Compared to the comprehensive baseline, this setting increases BD-rate by 162.77\% and decreases BD-PSNR by 4.731 dB. This performance is substantially worse than the 21.66\% BD-rate saving and 0.791 dB BD-PSNR gain achieved by our method.
		\begin{table} [!ht]
			\centering
			\renewcommand{\arraystretch}{1.6}
			\caption{Average BD-rate ($\%$) saving and BD-PSNR (dB) gain of related variants in the GGM compared to the baseline variant using COCO2017 datasets.} 
			\label{tab:Ablation}
			\scalebox{0.83}{
				\begin{tabular}{ccccc}  \hline
					\multicolumn{1}{c}{\textbf{Variants}} & \multicolumn{2}{c}{\textbf{PSNR}} & \multicolumn{2}{c}{\textbf{ROI-PSNR}}  \\ 
					\cmidrule(r){2-3} \cmidrule(l){4-5}
					& BD-rate $\downarrow$ & BD-PSNR $\uparrow$ & BD-rate $\downarrow$ & BD-PSNR $\uparrow$  \\ \hline
					\hspace{1em} Baseline & --  & --  & -- & --   \\ \hline
					\rowcolor{gray!20}
					Shape parameter $\mathbf{\beta}$ & & & & \\
					\hspace{1em} w $\textit{y = x}$ & 68.11  & -2.115  & 162.77  & -4.731   \\ 
					\rowcolor{gray!20}
					Scale parameter $\mathbf{\alpha}$ & & & & \\
					\hspace{1em} w $\textit{y = softplus(x)}$ & -58.92  & 1.726  & -15.93 & 0.588   \\ 
					\hspace{1em} w $\textit{y = x}$ & -55.01  & 1.605  & -19.52 & 0.709   \\ 
					\hspace{1em} w/o  $DLB$ & 9.68  & -0.264  & -3.92 & 0.146  \\ 
					\rowcolor{gray!20}
					$\epsilon$ values for $P(a,b)$ &  &  &  & \\ 
					\hspace{1em} w $\epsilon=1\times 10^{-7}$ & -24.71 & 0.680 & -18.19  & 0.651 \\  
					\hspace{1em} w $\epsilon=1\times 10^{-3}$ & -54.58 & 1.594 & -19.04  & 0.703 \\  \hline
					\hspace{1em} \textbf{Our} & \textbf{-56.45}  & \textbf{1.642} & \textbf{-21.66} & \textbf{0.791}  \\ \hline
			\end{tabular}}
		\end{table}
		We also visualize their RD curves in Fig.~\ref{fig:ablatdetseg}. When the shape parameter $\mathbf{\beta}$ is modeled with the linear function $\textit{y = x}$, it fails to accurately represent the latent distribution, leading to a substantial increase in overall bitrate and poor RD performance. Fig.~\ref{fig:betavisual} visualizes the feature maps of $\mathbf{\beta}$ produced by different functions. In these maps, deeper regions (larger $\mathbf{\beta}$ values) correspond to areas dominated by low-frequency content, while lighter regions (smaller $\mathbf{\beta}$ values) indicate the presence of high-frequency information. The $\mathbf{\beta}$ map generated by our \textit{Softplus} function accurately delineates this frequency distribution across the image. Benefiting from the properties of Softplus, the optimized $\mathbf{\beta}$ effectively identifies and represents the sparse high-frequency details present even in background regions. In contrast, the $\mathbf{\beta}$ map produced by the linear function $y=x$ is inaccurate, which severely undermines the GGM's ability to model latent distributions precisely.
		
		\textit{\textbf{Analysis of setting scale parameter $\mathbf{\alpha}$}}.
		To evaluate the impact of the scale parameter $\mathbf{\alpha}$, we compare different functions quantitatively (Table~\ref{tab:Ablation}) and RD curves (Fig.~\ref{fig:ablatdetseg}). Although the linear function $y=x$ (with a lower bound of 0.11) and the \textit{Softplus} function achieve competitive global PSNR, their ROI performance is substantially inferior to our \textit{Huber-like} function. Specifically, compared to our function, they achieve 2.14\% and 5.73\% lower BD-rate savings, respectively. Moreover, images reconstructed using these alternative functions lead to significantly degraded performance on downstream MV tasks. Fig.~\ref{fig:alphavisual} visualizes feature maps of $\mathbf{\alpha}$. Deeper regions (larger $\mathbf{\alpha}$) correspond to areas with high-frequency information, while lighter regions (smaller $\mathbf{\alpha}$) indicate low-frequency content. The \textit{Softplus} function tends to over-emphasize very small $\mathbf{\alpha}$ values, especially in background regions. This bias impairs accurate distribution modeling in critical areas and exacerbates train-test mismatch, degrading overall performance. Although the linear function $y=x$ prevents excessively small values via a fixed lower bound, it also imposes a strict constraint on the dynamic range of $\mathbf{\alpha}$. Consequently, $\mathbf{\alpha}$ cannot flexibly adapt to model the true latent distribution.
		
		In contrast, our \textit{Huber-like} function effectively mitigates both issues. It avoids the train-test mismatch associated with extremely small values without restricting the optimization range of $\mathbf{\alpha}$ parameters. By enabling a more accurate fit to the latent variables distribution, it consistently delivers superior coding performance across all evaluation metrics.

		\textit{\textbf{Analysis of dynamic lower bound setting for $\mathbf{\alpha}$}}. To quantify the impact on train-test mismatch, we apply different lower bounds to the scale parameter $\mathbf{\alpha}$ and calculate rate estimation errors $\Delta R$. As shown in Fig.~\ref{fig:alphas_r}, a smaller lower bound (e.g., 0.01) leads to a more pronounced $\Delta R$. The \textit{Softplus} function and the case without dynamic lower bound (“w/o DLB”) also exhibit significant $\Delta R$ when $\mathbf{\alpha}$ falls below approximately 0.11, exacerbating the mismatch problem.
		Fig.~\ref{fig:alphabetavisual} illustrates the distribution of $\mathbf{\alpha}$ values. Without the DLB constraint, $\mathbf{\alpha}$ values drift downward, with minimal values clustering in regions of larger shape parameter $\mathbf{\beta}$, which intensifies train-test mismatch. In contrast, applying our DLB maintains $\mathbf{\alpha}$ consistently above 0.11 across the full range of $\mathbf{\beta}$, effectively mitigating the mismatch.
		The performance impact is summarized in Table~\ref{tab:Ablation} and Fig.~\ref{fig:ablatdetseg}. The absence of the DLB substantially degrades compression performance. Compared to our full method, the “w/o DLB” variant reduces the BD-PSNR gain by 1.906 dB for the global image and by 0.645 dB for the ROI. It also yields lower accuracy on downstream MV tasks.
		
		\begin{figure}[!t] 
			\centering    
			\includegraphics[scale=0.27]{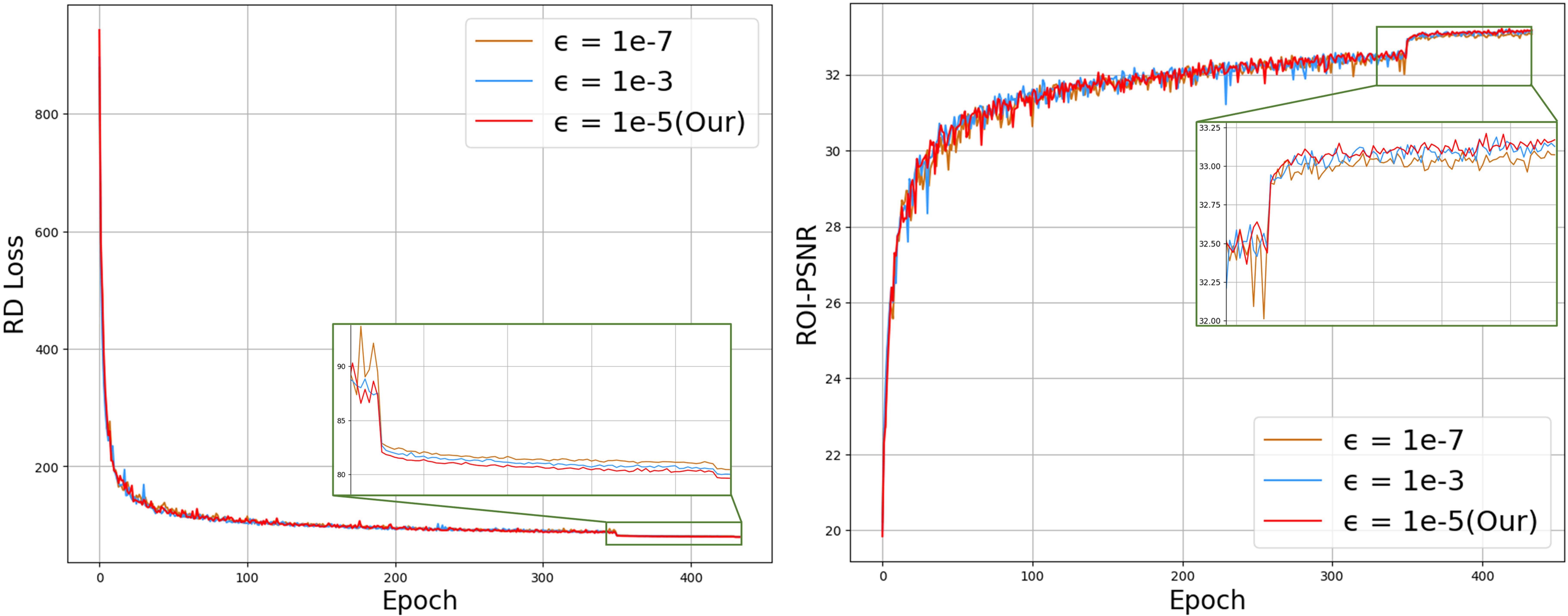} 
			\caption{The impact of $\epsilon$ values on compression network optimization and coding performance.}
			\label{fig:epsilon}
			\vspace{-0.15cm}
		\end{figure}
		\textit{\textbf{Analysis of setting $\epsilon$ value in function $P(a,z)$.}} We evaluate different values of the step size $\epsilon$ (Table~\ref{tab:Ablation}, Figs.~\ref{fig:ablatdetseg} and~\ref{fig:epsilon}). The results show that both excessively large and small $\epsilon$ degrade RD performance. A large $\epsilon$ (e.g., $1\times10^{-3}$) yields inaccurate gradients, preventing the GGM from modeling the latent distribution accurately. Conversely, a very small $\epsilon$ (e.g., $1\times10^{-7}$) introduces numerical instability and noise into the gradient computation, which also severely impairs performance. An appropriately chosen $\epsilon$ ensures an accurate and stable gradient approximation for $P(a,b)$, enabling the GGM to effectively capture the sharp-peaked and heavy-tailed characteristics of the latent features. This, in turn, enhances both ROI reconstruction quality and the accuracy of MV tasks.
		
		In summary, our design choices for the shape $\mathbf{\beta}$ (via a $\textit{Softplus}$ function) and the scale $\mathbf{\alpha}$ (via a $\textit{Huber-like}$ function and a DLB) are pivotal. Together, they enable accurate latent distribution modeling, mitigate train-test mismatch, and ultimately improve compression performance.

		\section{Conclusion}
		\label{6}
		This paper presents a unified RDO theoretical paradigm for ROI-based image compression. To overcome the GM's limitation in modeling the sharp-peaked and heavy-tailed distributions characteristic of ROI coding, we propose an elaborate GGM based on this theoretical paradigm, which has the learned shape $\mathbf{\beta}$ and scale $\mathbf{\alpha}$. We design specialized activation functions and a dynamic lower bound for $\mathbf{\alpha}$ to ensure stable optimization. Additionally, the finite central-difference scheme is validated as an efficient and sufficiently stable alternative to address the complex numerical integration problem in gradient computation of the $P(a,b)$ function within the GGM. Experimental results on multiple datasets show that our method reconstructs high-fidelity ROIs and achieves optimal performance in both compression and downstream MV tasks. Compared to classical probability models, our GGM yields significant gains in BD-rate and BD-PSNR, demonstrating superior coding efficiency. Future work will explore extending the GGM  to other image compression tasks.
		
		\bibliographystyle{IEEEtran}
		\bibliography{reference}

		\appendix
		\section*{Unified ROI-based Image Compression Paradigm with Generalized Gaussian Model}
		

		\subsection{Classic Generalized Gaussian Model}
		\label{appsub1}
\begin{figure}[!t] 
			\centering    
			\includegraphics[scale=0.28]{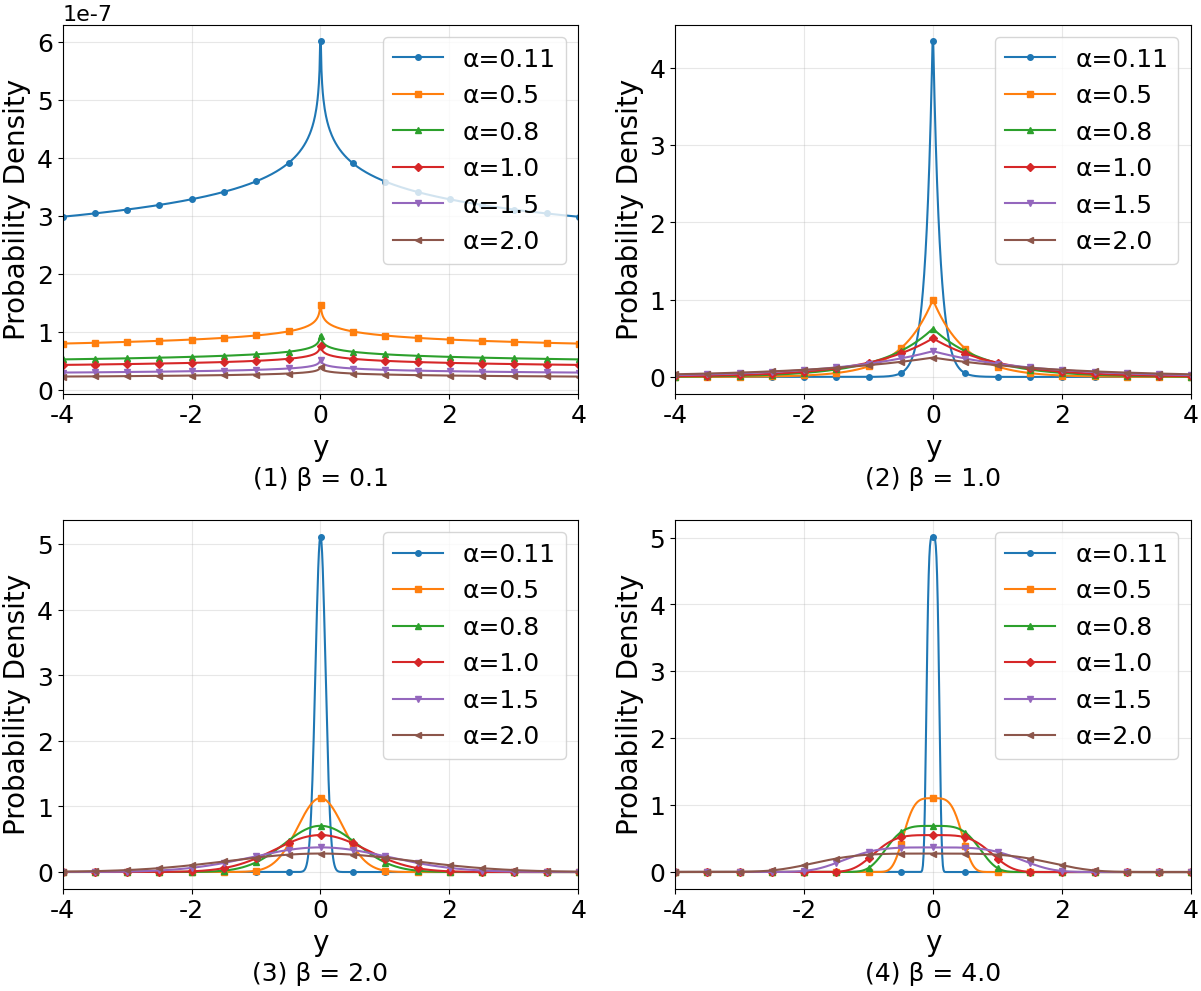} 
			\caption{The relationship between the PDF shape and scale parameters $\mathbf{\alpha}$ of GGM under different shape parameters $\mathbf{\beta}$.}
			\label{fig:ggmpdf}
		\end{figure}

		We visualize the influence of the scale and shape parameters in the GGM on modeling the latent space distribution, as shown in Fig.\ref{fig:ggmpdf}. It can be observed that the scale parameter $\mathbf{\alpha}$ controls the width or dispersion of the distribution, while the shape parameter $\mathbf{\beta}$ governs the kurtosis of the distribution—that is, its ``peakedness'' and ``tail thickness''.
		
		\subsection{Error Bound Analysis for Central Difference Scheme}
		\label{appsub2}
		To enable gradient-based stable optimization, we propose an efficient central difference scheme to approximate the gradient of $\gamma(a, b)$, thereby restoring differentiability during training. 
		\begin{equation}
			\begin{aligned}
				\frac{\partial}{\partial a}\gamma(a,b) \approx \frac{\gamma(a+\epsilon,b) - \gamma(a-\epsilon,b)}{2\epsilon},
			\end{aligned}
			\label{eqgamma1}
		\end{equation}
		The step size is $\epsilon = 1 \times 10^{-5}$. To analyze the error size of the central difference scheme used in replacing numerical integrals, we use Taylor expansion to derive Eq. \eqref{eqgamma1}. According to Taylor's theorem, it is known that
		\begin{equation}
			\begin{aligned}
				& f(a + \epsilon) = f(a) + f'(a)\epsilon + \frac{f''(a)}{2!}\epsilon^2 + O(\epsilon^3),\\
				& f(a - \epsilon) = f(a) - f'(a)\epsilon + \frac{f''(a)}{2!}\epsilon^2 - O(\epsilon^3)
			\end{aligned}
		\end{equation}
		Subtracting gives the central difference:
		\begin{equation}
			\begin{aligned}
				\frac{f(a + \epsilon) - f(a - \epsilon)}{2\epsilon} = f'(a) + O(\epsilon^2)
			\end{aligned}
		\end{equation}
		From this, it can be seen that the truncation error of the central-difference scheme is \(O(\epsilon^2)\), whereas that of the standard forward difference is \(O(\epsilon)\). Our scheme thus achieves a substantially more accurate approximation. The resulting error is controllable and meets the precision necessary for effective training.

		\subsection{Mask-guided Feature Enhancement Module}
		\label{appsub3}
\begin{figure}[!t] 
			\centering    
			\includegraphics[scale=0.54]{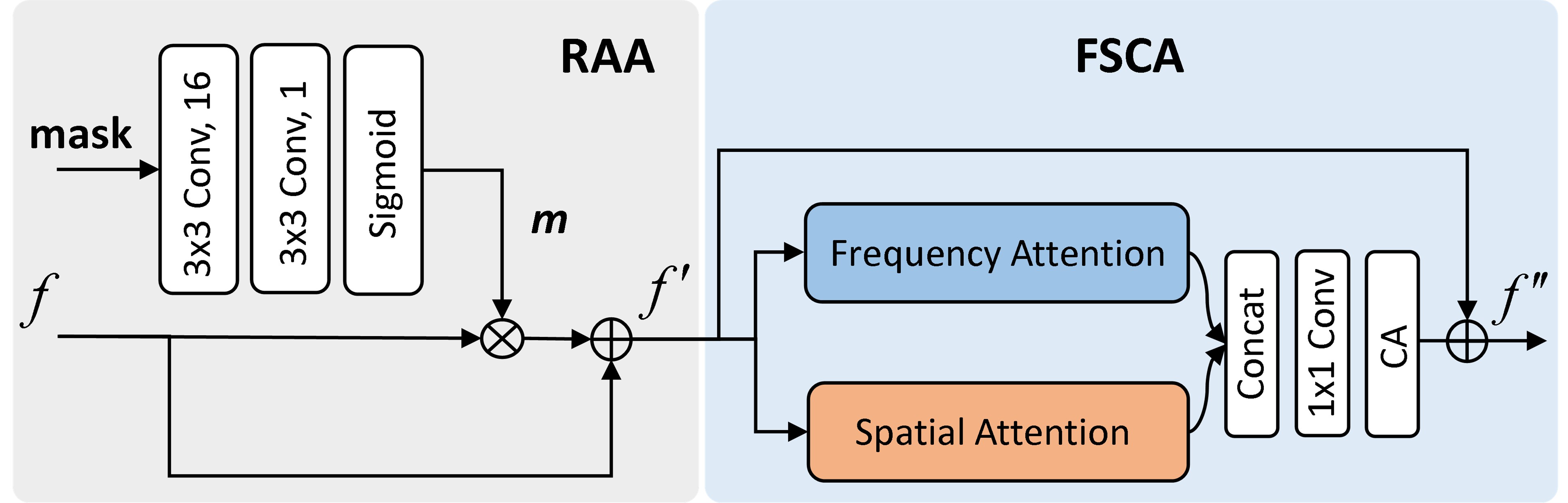} 
			\caption{Illustration of the MGFE Module. The MGFE Module is mainly composed of Region-Adaptive Attention (RAA) block and Fequency-Spatial Collaborative Attention (FSCA) block. Frequency
				attention (FA) branch and spatial attention (SA) branch in FSCA block}
			\label{fig:mgfe}
		\end{figure}

		To leverage the generated mask $\mathbf{M}$ for implicit bit allocation, MGFE modules are adopted, as shown in Fig.~\ref{fig:mgfe}, ${f}^{\prime\prime} = \text{MGFE}(\mathbf{M}, {f})$. The MGFE Module is mainly composed of a Region-Adaptive Attention (RAA) block and a Frequency-Spatial Collaborative Attention (FSCA) block. The RAA block is used as a transform function $\mathcal{F} (\cdot)$ to achieve important bit differentiation allocation. We integrate the MGFE module layer by layer in place of the initial Window Attention Module, which precisely distinguishes foreground from background regions, enabling high-quality region-adaptive coding. Formally, the MGFE module can be expressed in detail as
		\begin{equation}
			\begin{aligned}
				& {m} = \textit{Sigmoid}(\textit{Conv}_{3\times3}(\textit{Conv}_{3\times3}(\mathbf{M}))),\\
				& {f}^{\prime} = {f} \times {m} + {f}, \\
				&{f}_{att} = \text{Concat} \left(\textit{FreqAtt}({f}^{\prime}), \textit{SpatAtt}({f}^{\prime}) \right), \\
				&{f}^{\prime\prime} = {f}^{\prime} + \textit{CA} \left(\textit{Conv}_{1\times1} \left( {f}_{att} \right) \right).
			\end{aligned}
		\end{equation}
		
		In contrast to conventional spatial-only attention, the FSCA block integrates both frequency-domain and spatial-domain attention to collaboratively enhance key features across two orthogonal dimensions, thereby facilitating high-quality image representation. The FSCA block first takes the output features $\boldsymbol{f}^{\prime}$ of the RAA block as input. The input is then processed in parallel through both the frequency attention (FA) branch $FreqAtt(\cdot)$ and the spatial attention branch (SA) $SpatAtt(\cdot)$. The outputs from these two branches are concatenated (Concat) along the channel dimension. 
		A 1$\times$1 convolutional layer subsequently performs feature fusion and dimensionality reduction on the concatenated results. The fused features are then fed into a channel attention (CA) block to adaptively recalibrate the weights of different feature channels. Finally, attention features are added to the input $\boldsymbol{f}^{\prime}$ of the FSCA block through a residual connection to obtain the final feature maps $\boldsymbol{f}^{\prime\prime}$.
		
		\subsection{Compression Performance across HRSOD Dataset}
\begin{figure}[!t] 
			\centering    
			\includegraphics[scale=0.255]{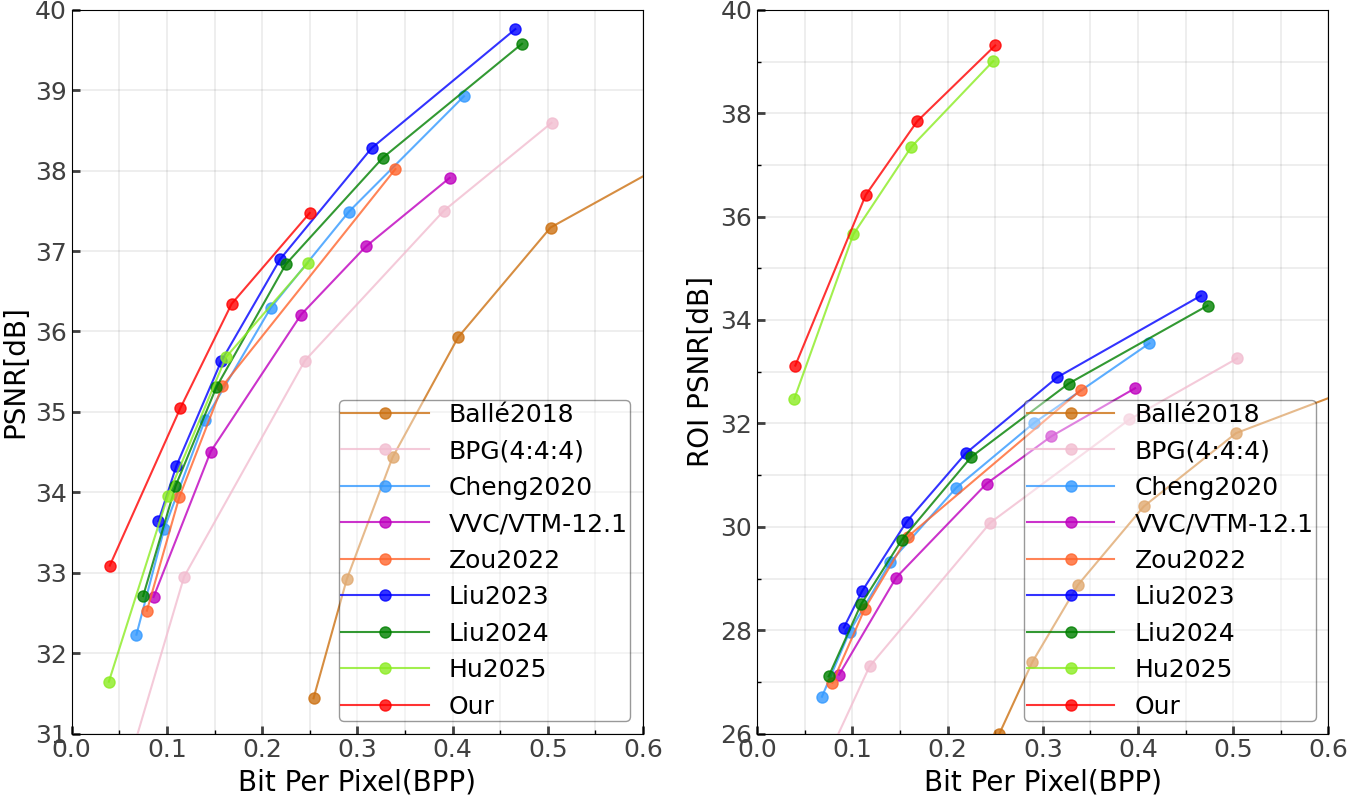} 
			\caption{RD performance averaged in terms of PSNR and ROI-PSNR across HRSOD dataset.}
			\label{fig:hrsodpsnr}
		\end{figure}

		We introduce an HRSOD dataset, which contains high-resolution images, to evaluate their RD performance. As shown in Fig.~\ref{fig:hrsodpsnr}, we plot the RD curves for different compression methods. Our method achieves the best overall compression performance in the ROI and non-ROI and outperforms the latest SOTA method, Hu2025. This further confirms the superiority and generalization capability of our GGM for ROI-based compression tasks.
		
		\subsection{Comparison of Different Probability Distribution Models}
\begin{figure}[!t] 
			\centering    
			\includegraphics[scale=0.25]{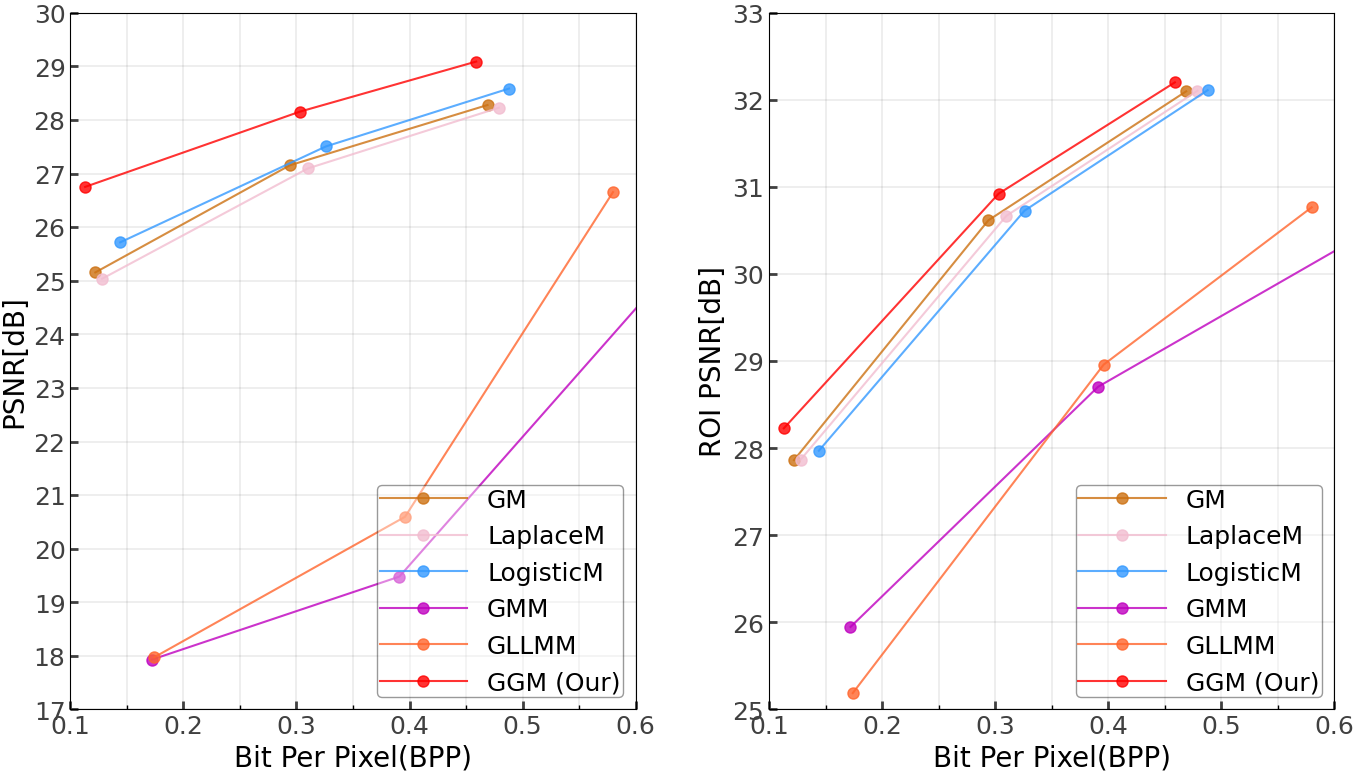} 
			\caption{RD performance using different PDFs.}
			\label{fig:diffRD}
		\end{figure}

		To analyze the impact of different probability distribution models (PDMs) on coding performance, we conduct a comparative experiment by replacing the GGM in our compression network with several classic PDMs. As shown in Fig.~\ref{fig:diffRD}, we plot the RD curves for different PDMs. Our GGM, with learned shape parameter $\mathbf{\beta}$, enables accurate modeling of the latent tensors distribution, achieving optimal coding performance for both the entire image and the ROI. Using GMM and GLLMM is unsuitable for modeling their true distribution and prone to causing model redundancy and overfitting.

		\vfill
		
	\end{document}